\documentclass[10pt]{article}

\usepackage{latexsym,amsfonts,amssymb}
\usepackage{color}
\usepackage{graphicx}
\usepackage{subfigure}
\usepackage{epsfig}
\usepackage{amsmath, amssymb}
\usepackage{fancybox}
\usepackage{listings}
\usepackage{verbatim}
\usepackage{url}
\usepackage{esint}
\usepackage{fancyhdr}
\usepackage{booktabs}
\usepackage{caption}
\usepackage{dsfont}
\input{epsf.sty}
\newtheorem{remark}{Remark}[section]
\newtheorem{example}{Example}[section]
\newtheorem{proposition}{Proposition}[section]

\newcommand{\sgn}{\mathop{\rm sgn}}
\newcommand{\diag}{\mathop{\rm diag}}

\newcommand{\remove}[1]{}

\title{{\bf Applications of the multi-sigmoidal deterministic and stochastic logistic models for plant dynamics}}

\author{
Antonio {\bf Di Crescenzo}\footnote{Corresponding author -- Email: adicrescenzo@unisa.it  -- Orcid: 0000-0003-4751-7341}
$^{,(1)}$
\
Paola {\bf Paraggio}\footnote{Email: pparaggio@unisa.it \ -- \ Orcid: 0000-0002-3308-7937}
$^{,(1)}$
\\
Patricia {\bf Rom\'an-Rom\'an}\footnote{Email: proman@ugr.es \ -- \ Orcid: 0000-0001-7752-8290}
$^{,(2)}$
\
Francisco {\bf Torres-Ruiz}\footnote{Email: fdeasis@ugr.es \ -- \ Orcid:  0000-0001-6254-2209}
$^{,(2)}$
\\
\\
\small (1) Dipartimento di Matematica, Universit\`a degli Studi di Salerno\\
\small Via Giovanni Paolo II n.\ 132, I-84084 Fisciano (SA), Italy.
\\
\\
\small (2) Department of Statistics and Operations Research\\
\small Faculty of Sciences, University of Granada, Avenida Fuente Nueva s/n, 18071  Granada, Spain;  \\
\small Institute of Mathematics of the University of Granada (IEMath-GR)\\
\small Calle Ventanilla, 11, 18001, Granada, Spain.
}
\date{}

\begin{document}

\maketitle

%
\begin{abstract}
We consider a generalization of the classical logistic growth model introducing more than one inflection point. The growth, called multi-sigmoidal, is firstly analyzed from a deterministic point of view in order to obtain the main properties of the curve, such as the limit behavior, the inflection points and the threshold-crossing-time through a fixed boundary. We also present an
application in population dynamics of plants based on  real data.
Then, we define two different birth-death processes, one with linear birth and death rates and the other with quadratic rates, and we analyze their main features. The conditions under which the processes have a mean of multi-sigmoidal logistic type and the first-passage-time problem are also discussed. Finally, with the aim of obtaining a more manageable stochastic description of the growth, we perform a scaling procedure leading to a lognormal diffusion process with mean of multi-sigmoidal logistic type. We finally conduct a detailed probabilistic analysis of this process.



\end{abstract}

	
\section{Introduction}
The logistic model is a sigmoidal growth model characterized by an initial slow growth followed by an explosion of exponential-type which flattens up to an equilibrium status (known as carrying capacity).
It is a growth curve particularly useful to describe evolution phenomena in restricted environments. In literature, there are many growth models with an S-shape, such as Gompertz, Korf or mixed models (see Brauer and Castillo-Chavez \cite{Brauer2010}).
The applications of sigmoidal curves are various and they involve several contexts of interest which go from biology to medicine, from ecology to software reliability.
For example, in the recent works of Rajasekar \textit{et al.} \cite{Rajasekaretal2020a,Rajasekaretal2020b}, the authors analyze a stochastic version of SIR model for the diffusion of the COVID-19 pandemic, by supposing that the number of susceptible individuals follows a logistic-kind growth.
Moreover, regarding software reliability, in their recent work Erto and Lepore \cite{Erto2020} define a new kind of S-shaped curve which, under suitable choices of the involved parameters, has got more than one inflection point. Indeed, it is possible that a population reaches its limit value after various successive steps. This is the reason why recent investigations address their interest to a generalization of the sigmoidal models by the introduction of multiple inflection points. Such generalizations are the so-called multi-sigmoidal models
(see for example Rom\'an-Rom\'an {\em et al.}\ \cite{Romanetal2019}).
The multi-sigmoidal logistic model, in particular, is appropriate to describe the maturation of some fruit species (such as peaches or coffee berries) which shows a trend with multiple fluctuations (see, for example Fernandes \textit{et al.}\ \cite{Fernandesetal2017}). Another different application of the multi-sigmoidal model can be found in Cairns \textit{et al.}\ \cite{Cairnsetal2008}, where a double sigmoidal fitting is considered to compare fatigue profiles obtained with different stimulation protocols in isolated slow-twitch soleus and fast-twitch extensor digitorum longus (EDL) muscles of mice.
Furthermore, in the study of energy resources, in particular oil, there are models associated with logistic growth, such as the Hubbert model,
which is used to determine the peak of oil production. However, in recent years it has been observed how the behavior of oil production
shows various peaks, which is related to the presence of various inflection points in the underlying logistics models (see
Maggio and Cacciola \cite{MaggioCacciola2009}, and Saraiva \textit{et al.}\ \cite{Saraiva_etal2014}).
\par
All the afore-mentioned curves are deterministic, in the sense that they are described by precise differential equations
({see, for example Section 2.2 of Banks \cite{Banks1994}). Even if they are useful models, they do not take into account random fluctuations which characterize the real world. For this reason, many efforts have been realized in order to introduce dynamic models related to these curves. Among them, stochastic diffusion processes stand out. These processes are governed by a stochastic differential equation obtained by adding to the deterministic equation a noise term which is represented, most of the times, by a Wiener process (see, as a reference, \O ksendal \cite{Oksendal2003}). They are constructed in such a way that their mean is equal to the growth curve under analysis.
The choice of the noise type which is added to the deterministic equation depends on the context.
For instance, Scholman \cite{Schloman2018} suggests to modify the differential equation which describes the logistic growth, by adding a random term represented by a Poisson process, this being more suitable for the description of growth phenomena with random catastrophes.  Otherwise, other investigations propose to introduce the random environment by defining particular birth-death processes with a mean identical to the growth curve (see Di Crescenzo and Paraggio \cite{DiCrescenzoParaggio2019}, Di Crescenzo and Spina \cite{DiCrescenzoSpina2016} and Ricciardi \cite{Ricciardi1986}). Other diffusion approximations of the logistic growth have been performed by Campillo {\em et al.} in \cite{Campillo2016} where the author studies the corresponding stochastic growth model with extinction, by Kink in \cite{Kink2018}, by Nobile and Ricciardi in \cite{NobileRicciardi1984a,NobileRicciardi1984b}. See also Di Crescenzo {\em et al.}\ \cite{DiCrescenzoetal2016}
for a method  to construct tractable diffusion processes  suitable for describing populations subject to rapid growth.
\par
The need of constructing stochastic processes whose mean follows a given trend emerges in several applications in which
the intrinsic random fluctuations cannot be neglected and require the construction of appropriate random dynamic systems.
Hence, stimulated by the above mentioned research lines, in the present paper we dedicate attention to both the described strategies.
Indeed, we define a linear and time-inhomogeneous birth-death process and a diffusion process,  both processes possessing a mean of multi-sigmoidal logistic type. We also address our attention to a particular birth-death process with quadratic rates from which we derive the diffusion process, as the limit under a suitable scaling.  In literature, the analysis of quadratic birth-death processes is a quite hard task and thus not amply discussed. In particular, Lenin and Parthasarathy \cite{LeninParthasarathy2000} and Parthasarathy and Vijayashree \cite{ParthasarathyVijayashree2002}  studied the Markovian queues with finite capacity  in which the arrivals and the service completions are governed by quadratic functions. The probability generating function has been analyzed using suitable properties of tridiagonal matrices. The probabilities of quadratic birth-death processes can be also determined by means of Laplace transform as done by Lenin and Parathasarathy in \cite{LeninParthasarathy1997,LeninParthasarathy1998}.  We follow the approach of Letessier and Valent \cite{LetessierValent1984}, Roehner and Valent \cite{RoehnerValent1982} and Valent \cite{Valent1996} and thus we study the probability generating function deriving from it a differential equation for the mean.
We also consider the first-passage-time (FPT) problem both for the birth-death process with linear rates and for the approximating diffusion process through special boundaries as done in Giorno and Nobile \cite{GiornoNobile2019} and Guti\'errez {\em et al.} \cite{Gutierrez1997}.
\par
Let us describe the contents of the paper. In Section \ref{Section2} the multi-sigmoidal logistic curve is defined and its main features, such as the limit behavior and the inflection points, are described. The threshold crossing time problem is also discussed. In Section \ref{section3} we consider a real data set concerning the maturation of coffee fruits and an approximation of multi-sigmoidal logistic type is performed. The choice of the best fit is based on the minimization of the cumulative square error. Section \ref{section4} is devoted to the introduction of the linear birth-death process. We find a sufficient and necessary condition to have a mean of multi-sigmoidal logistic type and we analyzed the FPT problem through a fixed boundary. The quadratic birth-death process is introduced in Section \ref{Section5}: we study the probability generating function and the corresponding partial differential equation from which we obtain an ordinary differential equation solved by the mean of the process. After a discussion regarding the asymptotic behavior of the process, we perform a scaling with diffusive approximation that leads the birth-death process to a lognormal diffusion one. In Section \ref{Section6} we study its main properties and the corresponding FPT problem. We also point out that both the unconditional and conditional mean of the aforementioned diffusion process are of multi-sigmoidal logistic type.

\section {The multi-sigmoidal logistic function}\label{Section2}
The classical logistic equation is expressed as
 %
$$
 \frac{d}{dt} l(t)=rl(t)\left[1-\frac{\eta }{C}l(t)\right],\qquad t\ge t_0,
$$
where $r>0$ is the  intrinsic growth rate, and ${\eta/C}>0$. In more general instances the rate $r$ may be taken as time-varying.
When $r$ is replaced by a polynomial $P(t)$, the solution of the corresponding ordinary differential equation (ODE), with initial condition $l(t_0)=l_0>0$, is
$$
 l(t)=\frac{l_0 e^{Q(t)-Q(t_0)}}{1-\displaystyle \frac{\eta}{C}l_0\left(1-e^{Q(t)-Q(t_0)}\right)},
 \qquad t\ge t_0,
$$
where $Q(t)-Q(t_0)=\int_{t_0}^t P(\tau) d\tau$. In this case, if $Q(t)\to +\infty$ for $t\to +\infty$, then the limit of $l(t)$
for $t\to +\infty$ is given by $\frac{C}{\eta}$, so that the carrying capacity is independent on the initial value $l_0$.
\par

Aiming to construct a similar generalization of the logistic growth model in which the carrying capacity depends on the initial value,
now we focus on the following equation:
 \begin{equation}\label{5}
 \frac{d}{dt} l_m(t)=h_\theta (t)l_m(t),\qquad t \ge t_0,
 \end{equation}
with
 \begin{equation}\label{htheta}
 h_\theta(t):=\frac{P_\beta(t)e^{-Q_\beta (t)}}{\eta + e^{-Q_\beta (t)}},
 \end{equation}
for $\eta >0$, $\theta=(\eta, \beta^T)^T$ with $\beta^T=(\beta_1,\dots, \beta_p)$, and where $Q_\beta$ and $P_\beta$ are polynomial defined as
 \begin{equation}\label{Qbeta}
 Q_\beta (t)=\sum_{i=1}^{p}\beta_i t^i, \qquad \beta_p>0
\end{equation}
and
\begin{equation}\label{PBeta}
P_\beta(t)=\frac{d}{dt}Q_\beta(t).
\end{equation}
Under the given assumptions, the solution of the ODE (\ref{5}) with initial condition $l_m(t_0)=l_0>0$ is given by
 \begin{equation}\label{lm}
 l_m(t) 
 =l_0\frac{\eta +e^{-Q_\beta (t_0)}}{\eta + e^{-Q_\beta (t)}},
 \qquad t\geq t_0.
\end{equation}
The curve given in (\ref{lm}) is named multi-sigmoidal logistic curve, since it exhibits various kinds of shapes
characterized by multiple inflection points. Indeed, assuming that $l_0$ and $t_0$ are fixed,
suitable choices of the parameters $\eta, \beta_1,\dots, \beta_p$ lead to a model with multiple fluctuations.
The approach based on the use of polynomials to construct
flexible growth curves has been successful exploited in \cite{Romanetal2019}
for a Gompertz-type model. Recalling the assumption $\beta_p>0$, one has the limit
 \begin{equation} \label{limit}
  \lim_{t\to \infty}l_m(t)
  =l_0 \frac{\eta +e^{-Q_\beta (t_0)}}{\eta}= \frac{C}{\eta},
\end{equation}
with $C=C(l_0,\eta,\beta, t_0)$, so that the carrying capacity $C/\eta$ depends on the relevant parameters.

Various characteristics of the curve (\ref{lm}) are provided in Table \ref{tab:Tabella1}.
Even if the curve (\ref{lm}) is positive and bounded, with initial value $l_0$ and limit value $C/\eta$, its
monotonicity intervals cannot be established in general, since they depend on the values of the parameters $\theta=(\eta, \beta^T)^T$.
Some plots of the multi-sigmoidal logistic function $l_m$ for special choices of the parameters are given in Figure \ref{fig:Figura1},
whereas  Figure \ref{fig:Figura2222} shows some plots of the derivative of $l_m$.
\begin{table}[t]
\caption{
Some features of the multi-sigmoidal logistic function where, for coinciseness, we   set
$Q_{\beta}^{(i)}(t)=Q_{\beta}(t)|_{\beta_i=0}=Q_{\beta}(t)-\beta_it^i$,
and $P_{\beta}^{(i)}(t)=P_{\beta}(t)|_{\beta_i=0}=P_{\beta}(t)-i \beta_it^{i-1}$.
} 
	\centering
	 \footnotesize
	\begin{tabular}{|c|c|c|c|c|c|c|} 
		\hline
		$\quad$		 	&$t\to \infty$ 	&$\eta\to 0$ 	&$\eta\to \infty$ &$\beta_i\to 0$		&$\beta_i\to \infty$ 		&$\beta_i\to -\infty,\; i\neq p$\\	
		\hline
		$l_m(t)$		& $C/\eta$		& $Ce^{Q_\beta(t)}$		&$0$ &$\displaystyle\frac{C}{\eta+e^{-Q_{\beta}^{(i)}(t)}}$ 	& $C/\eta$ 	&$0$	\\[0.4cm]
		$\frac{d }{dt}l_m(t)$		& $0$		& $CP_\beta(t)e^{Q_\beta(t)}$		&$0$ 	&$\displaystyle \frac{C P_{\beta}^{(i)}(t) e^{-Q_{\beta}^{(i)}(t)}}{\left(\eta+e^{-Q_{\beta}^{(i)}(t)}\right)^2}$ 	& $0$ &$0$	\\
	\hline	
	\end{tabular}	
\label{tab:Tabella1}	
\end{table}
%
\begin{figure}[t]
	\centering
	\hspace*{-1cm}
	\subfigure[]{\includegraphics[scale=0.5]{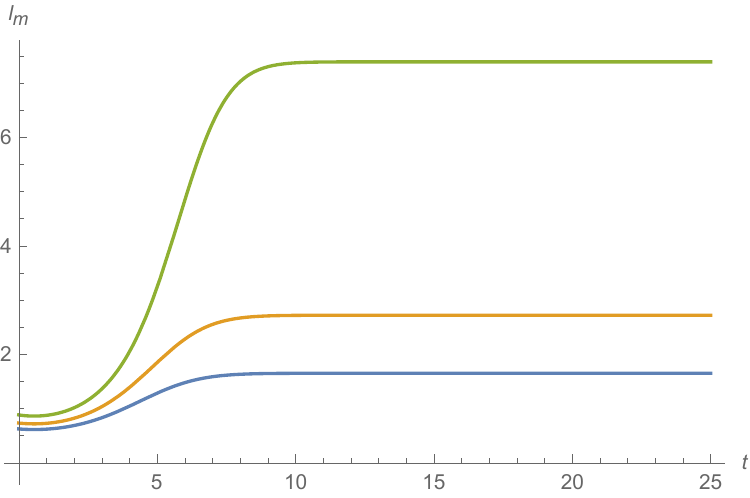}}\qquad
	\subfigure[]{\includegraphics[scale=0.5]{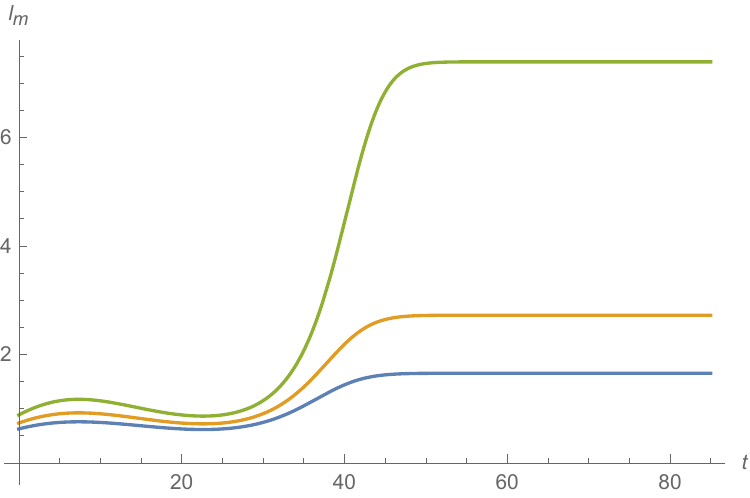}}
	\caption{The multi-sigmoidal logistic function for (a) $p=2$ and (b) $p=3$.
	The values of the parameters are $\eta=e^{-0.5}, e^{-1}, e^{-2}$ (from bottom to top), $l_0=(\eta+1)^{-1}$ and (a) $Q_\beta(t)=-0.1t+0.09t^2$, (b) $Q_\beta(t)=0.1t-0.009t^2+0.0002t^3$. }
	\label{fig:Figura1}
\end{figure}
%
	%
	\begin{figure}[t]
	\centering
	\hspace*{-1cm}
	\subfigure[]{\includegraphics[scale=0.5]{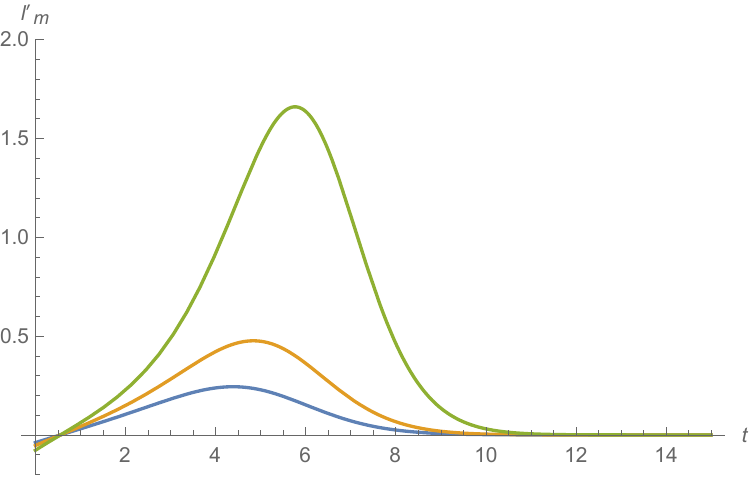}}\qquad
	\subfigure[]{\includegraphics[scale=0.5]{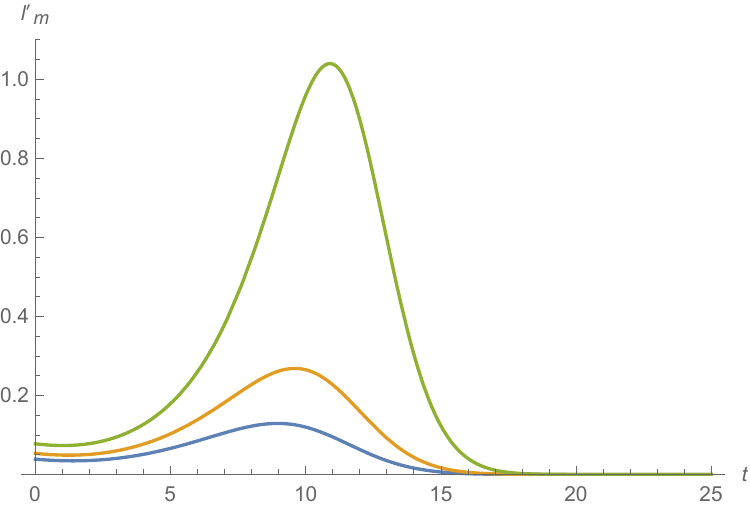}}\\
	\caption{The derivative of the multi-sigmoidal logistic function for (a) $p=2$ and (b) $p=3$.
	The values of the parameters are $\eta=e^{-0.5}$, $e^{-1}$, $e^{-2}$ (from bottom to top), $l_0=(\eta+1)^{-1}$
	and (a) $Q_\beta(t)=-0.1t+0.09t^2$, (b) $Q_\beta(t)=0.1t-0.009t^2+0.0002t^3$.}
	\label{fig:Figura2222}
	\end{figure}
 	%
\par

We remark that, due to Eq.\ (\ref{5}), the function given in (\ref{htheta}) plays the role of a time-dependent growth rate.
In general, specific choice of $h_\theta(t)$ allow to construct suitable growth models.
In this framework we mention the recent contributions by Asadi {\em et al.}\ \cite{Asadi_etal2020} and Chakraborty {\em et al.}\ \cite{Chakraborty_etal2019}.
It is easy to see that the function (\ref{htheta}) is continuous, bounded, and positive on the intervals in which $l_m$ is increasing.
Some plots of the function $h_\theta$ are given in Figure \ref{fig:Figura3333}.

  %
  \begin{figure}[t]
  	\centering
  	\hspace*{-1cm}
  	\subfigure[]{\includegraphics[scale=0.5]{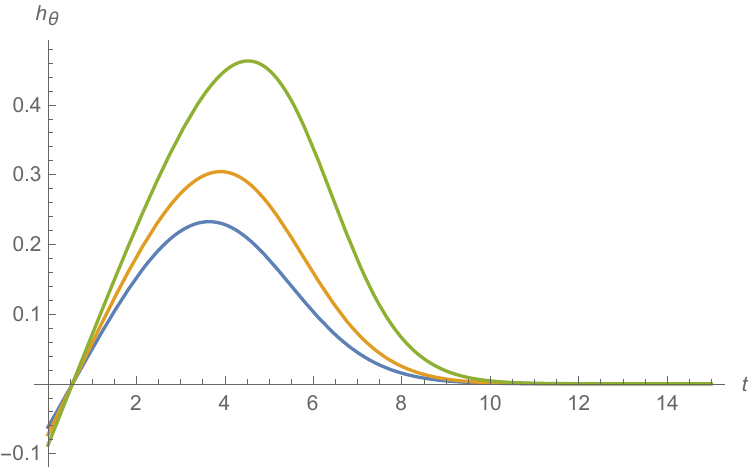}}\qquad
  	\subfigure[]{\includegraphics[scale=0.5]{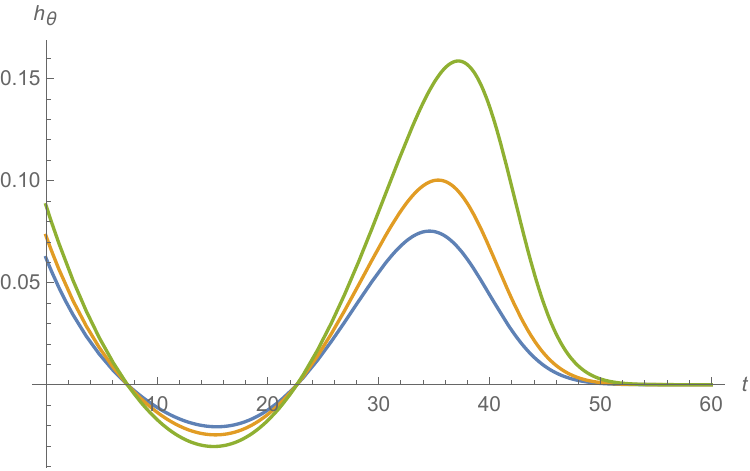}}\\
  	\caption{The function $h_\theta$ for (a) $p=2$ and (b) $p=3$.
	The values of the parameters are $\eta=e^{-0.5}$, $e^{-1}$, $e^{-2}$ (from bottom to top for large $t$) and (a) $Q_\beta(t)=-0.1t+0.09t^2$, (b) $Q_\beta(t)=0.1t-0.009t^2+0.0002t^3$.}
  	\label{fig:Figura3333}
  \end{figure}
  %
 %

\begin{remark}\label{rem:1}
The analysis of the  multi-sigmoidal logistic growth model \eqref{lm} can be performed, without loss of generality, by taking $t_0=0$.
Indeed, by setting $t'=t-t_0$ we obtain the similar model
$$
 l_m(t)|_{t_0=0} =l_0 \frac{ \eta+1}{ \eta + e^{-  Q_{\beta}(t)}}
  =l_0 \frac{\tilde \eta+e^{-\tilde Q_{\gamma}(0)}}{\tilde\eta + e^{-\tilde Q_{\gamma}(t')}}=:\tilde l_m(t'),
  \qquad t'\ge 0,
$$
since $Q_\beta(0)=0$, for   $\tilde\eta:=\eta e^{-\beta_0}$ and $\tilde Q_{\gamma}(t-t_0):=Q_{\beta}(t)+\beta_0$, where
$\beta_0=\sum_{k=1}^{p}\gamma_k(-t_0)^k$ and  the parameters $\gamma^T=(\gamma_1,\ldots,\gamma_p)$ can be obtained
from $\beta_i=\sum_{k=i}^{p}\binom{k}{i}\gamma_k(-t_0)^{k-i}$, $i=1,\dots,p$.
\end{remark}

\begin{remark}\label{rem:2}
With the purpose of obtaining a more flexible growth model, the multi-sigmoidal logistic model $l_m$ can be properly generalized to the case
in which one or more exponents of the polynomial $Q_\beta$ are rational or real.
In this way the application to real data leads to a better goodness-of-fit, as will be shown in Section \ref{section3}.
However, the number of parameters to estimate become larger and this can cause a higher computational cost.
\end{remark}

  \subsection{Inflection points}
  The investigation about the inflection points of a multi-sigmoidal logistic function is of great interest in applications, since, as we will see in Section \ref{section3}, some populations show a growth pattern of multi-sigmoidal type. Noting that
$$
 \frac{d^2}{d t^2}l_m(t)
 =\frac{l_0 e^{-Q_\beta(t)} \left(\eta+e^{-Q_\beta(t_0)}\right)}{(\eta+e^{-Q_\beta(t)})^3}
 \left[\frac{d}{dt}P_\beta(t)\left(\eta+e^{-Q_\beta(t)}\right)-P^2_\beta(t)\left(\eta-e^{-Q_\beta(t)}\right)\right],
$$
and recalling (\ref{PBeta}),
the inflection points solve the following  equation in the unknown $t\geq t_0$
  \begin{equation}\label{6}
  \frac{d^2}{dt^2}Q_\beta (t)=\left( \frac{d}{dt}Q_\beta (t)\right)^2 \frac{\eta- e^{-Q_\beta (t)}}{\eta+ e^{-Q_\beta (t)}},
  \end{equation}
where $Q_\beta$ is given in Eq.\ \eqref{Qbeta}.
 If the function $Q_\beta$ attains a minimum and a maximum for $t=t_1$ and $t=t_2$, respectively,
 then there exists $\tau\in[\min\{t_1, t_2\}, \max\{t_1, t_2\}]$ which solves eq.\ \eqref{6}.
   Hence, due to the transcendental nature of equation \eqref{6}, in general it is not possible to give an explicit expression of the inflection points, so that one is forced to adopt numerical methods.
  See Figure \ref{fig:Figura4} for some plots of the second derivative of the multi-sigmoidal logistic function.
  \par
  %
  \begin{figure}[t]
  	\centering
  	\hspace*{-1cm}
  	\subfigure[]{\includegraphics[scale=0.5]{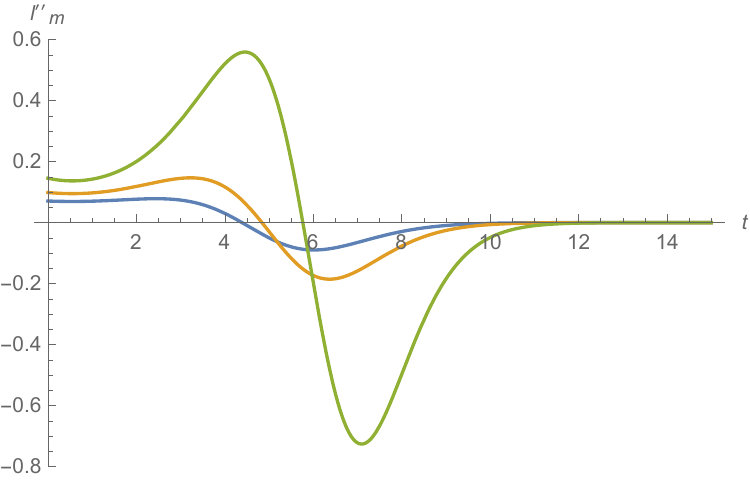}}\qquad
  	\subfigure[]{\includegraphics[scale=0.5]{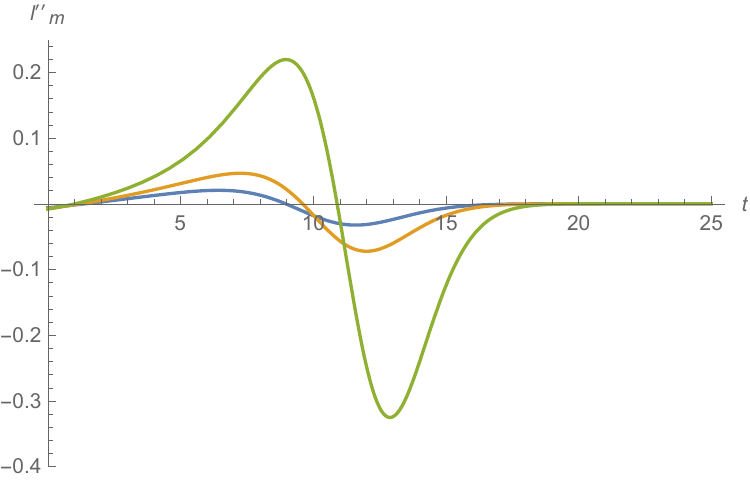}}\\
  	\caption{The function $l''_m$ for (a) $p=2$ and (b) $p=3$. The values of the parameters are
	$\eta=e^{-0.5}$, $e^{-1}$, $e^{-2}$ (from bottom to top near the origin), $l_0=(\eta+1)^{-1}$  and (a) $Q_\beta(t)=-0.1t+0.09t^2$, (b) $Q_\beta(t)=0.1t-0.009t^2+0.0002t^3$.}
  	\label{fig:Figura4}
  \end{figure}
  %
\par
  We can also analyze the curve in proximity of its inflection points by means of a linear approximation,
  as already done in \cite{DiCrescenzoParaggio2019} and \cite{DiCrescenzoSpina2016}.
  Considering an inflection point $t^*$, we denote by $\mu^*$ the so-called maximum specific growth rate defined as follows
  \begin{equation*}
  \mu^*=\left.\begin{matrix}
  \displaystyle\frac{d }{dt}l_m(t)
  \end{matrix}\right|_{t=t^*},
  \end{equation*}
  which represents the slope of the line tangent to the curve in the point $t^*$. Moreover, we denote by $\lambda^*$ the corresponding lag time, that is defined as the intersection between the $x$-axis and the above mentioned tangent.

We recall that the lag time provides useful information for the description of the growing processes that exhibit lag, growth, and asymptotic phases
(see, for instance,  Zwietering {\em et al.}\ \cite{Zwietering_etal}).
Indeed, the lag time corresponds to the initial time that would allow an ideal population, growing with constant maximum rate,
to reach the same size of the previous population at the inflection point.

Taking into account the expression of the derivative of the curve $l_m$, it follows that
  \begin{equation*}
  \mu^*=\frac{l_0 \left( \eta+ e^{-Q_\beta(t_0)} \right) e^{-Q_\beta(t^*)}P_\beta(t^*)}{\left(\eta+e^{-Q_\beta(t^*)}\right)^2},\qquad \lambda^*=t^*-\frac{1}{h_\theta (t^*)},
  \end{equation*}
  where the function $h_\theta$ has the expression given in \eqref{htheta}. Note that $\mu^*$ can be either positive or negative according to the sign of $P_\beta(t^*)$. If $\beta_i\to\pm\infty$, for $i=1,\dots,p-1$ or $\beta_p\to+\infty$ or $\eta\to+\infty$, then $\mu^*$ tends to $0$ and thus the tangent line tends to be parallel to the $x$-axis. Indeed, in these limit cases the curve
$l_m$ degenerates into a horizontal line (see Table \ref{tab:Tabella1}).

  \subsection{Threshold crossing problem}\label{TCP}
  In this section, we aim to analyze the time that a population modeled by the multi-sigmoidal logistic function \eqref{limit} spends below (or above) an upper (or lower) constant threshold. These boundaries may represent critical values related to the dynamics of the modeled population evolution.
  In both cases, the thresholds are taken as a function of the initial value $l_0$, since in various applications it is interesting to investigate
  the first  time when the population reaches a specific  depending on the known initial value.

  \par
  Considering an upper boundary $B_U$ with $B_U>l_0>0$, we can express it as a multiple of the initial value $l_0$, that is
  \begin{equation*}
  B_U= n l_0,\qquad n>1.
  \end{equation*}
  We define $\theta_U$ as the first time instant in which the function $l_m$ crosses the boundary $B_U$, that is
  \begin{equation*}
  \theta_U=\min\left\{t\ge t_0 \colon  l_m(t)=B_U\right\}.
  \end{equation*}
Clearly, for some choices of the parameters $\beta_i$ the set $\left\{t\ge t_0 \colon  l_m(t)=B_U\right\}$ may be empty. In this case, we set $\theta_U=+\infty$. Otherwise, from the definition of $\theta_U$, it immediately follows that $l_m(\theta_U)=B_U=n l_0$ and thus
  \begin{equation*}\label{cond1}
  -Q_\beta(\theta_U)=\log\left[\frac{\eta+e^{-Q_\beta(t_0)}}{n}-\eta\right],
  \end{equation*}
  with $(1-n)\eta+e^{-Q_\beta(t_0)}>0$ and $n>1$.
  \par
  Since the function $l_m$ may be decreasing in some intervals, we can analyze the time spent by the function $l_m$ above a lower boundary $B_L$ with $0<B_L<l_0$.
  In analogy with the previous case, the threshold $B_L$ can be expressed as a submultiple of the initial value $l_0$, i.e.
  \begin{equation*}
   B_L=\frac{1}{n}l_0,\qquad n>1.
  \end{equation*}
  We denote by $\theta_L$ the first time instant in which the function $l_m$ passes through the lower boundary $B_L$, that is
  \begin{equation*}
  \theta_L=\min\left\{t\ge t_0 \colon l_m(t)=B_L\right\}.
  \end{equation*}
 Also in this case, when the set $\left\{t\ge t_0\colon l_m(t)=B_L\right\}$ is empty, we consider $\theta_L=+\infty$,
 otherwise, when the set $\left\{t\ge t_0\colon l_m(t)=B_L\right\}$ is not empty, it easily follows from the definition of
 $\theta_L$ that $l_m(\theta_L)=B_L=\frac{1}{n}l_0$ and thus
 \begin{equation*}\label{cond2}
 -Q_\beta(\theta_L)=\log\left[n\left({\eta+e^{-Q_\beta(t_0)}}\right)-\eta\right],
 \end{equation*}
 with $n>1$.
  The threshold crossing times $\theta_U$ and $\theta_L$ are plotted in Figure \ref{fig:Figura10} for some choices of the parameters.
  \begin{figure}[t]
  	\centering
  	\hspace*{-1cm}
  	\subfigure[]{\includegraphics[scale=0.5]{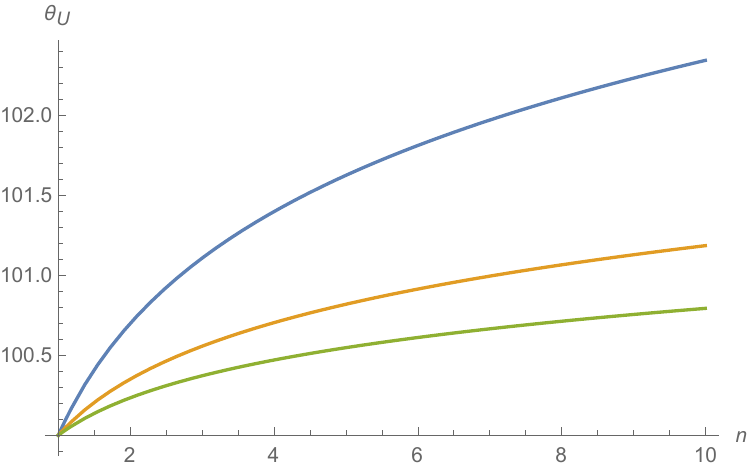}}\qquad
  	\subfigure[]{\includegraphics[scale=0.5]{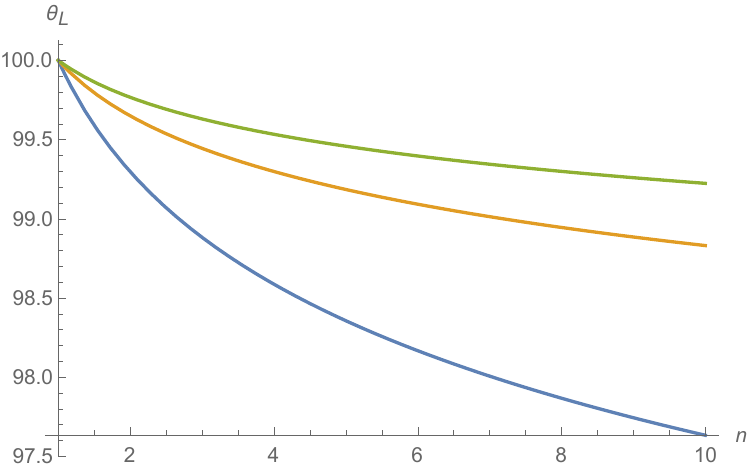}}
  	\caption{The threshold crossing times for $\eta=0.01$ and (a) $\theta_U$, for
	$Q_\beta(t)=-t+0.01t^2$, $Q_\beta(t)=-2t+0.02t^2$, $Q_\beta(t)=-3t+0.03t^2$ (for top to bottom),
	 and (b) $\theta_L$  (for the same choices of  $Q_\beta(t)$, in reversed order).}
  	\label{fig:Figura10}
  \end{figure}
%
\section{An application }\label{section3}
In this section we consider an application of the previous results in the context of
population dynamics of plants.
The analysis involves real data for which the multi-sigmoidal logistic function \eqref{lm} provides a good fit.
Specifically, we determine the values of the parameters involved in the definition of \eqref{lm} in order to minimize the square error.
\par
It is well known that some fruits show a growth with a multi-sigmoidal pattern, for this reason their development can be modeled by a particular multi-sigmoidal logistic curve. The data provided in Figure \ref{fig:Figura8} are taken from da Cuhna and Volpe \cite{CuhnaVolpe2011} and are concerning the accumulated fresh mass of coffee berries
Obat$\tilde{\rm a}$ IAC $1669-20$. Since the development of coffee fruits depends on solar radiation, the authors consider particular positions of the plants based on the apparent trajectory of the sun and we refer to the alignment $51^{\rm o}-231^{\rm o}$ (azimuth).
\begin{figure}[t]
	\centering
	\hspace*{-1cm}
	\subfigure[]{\includegraphics[scale=0.4]{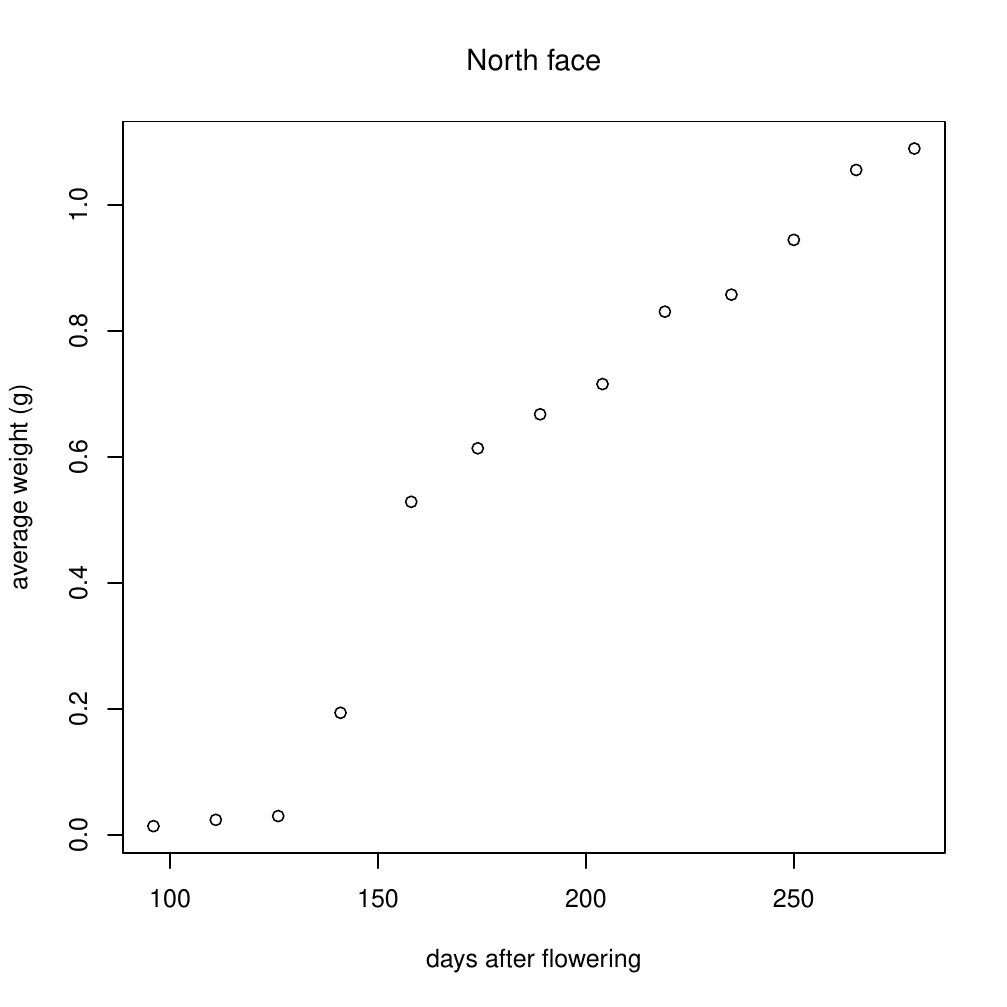}}\qquad
	\subfigure[]{\includegraphics[scale=0.4]{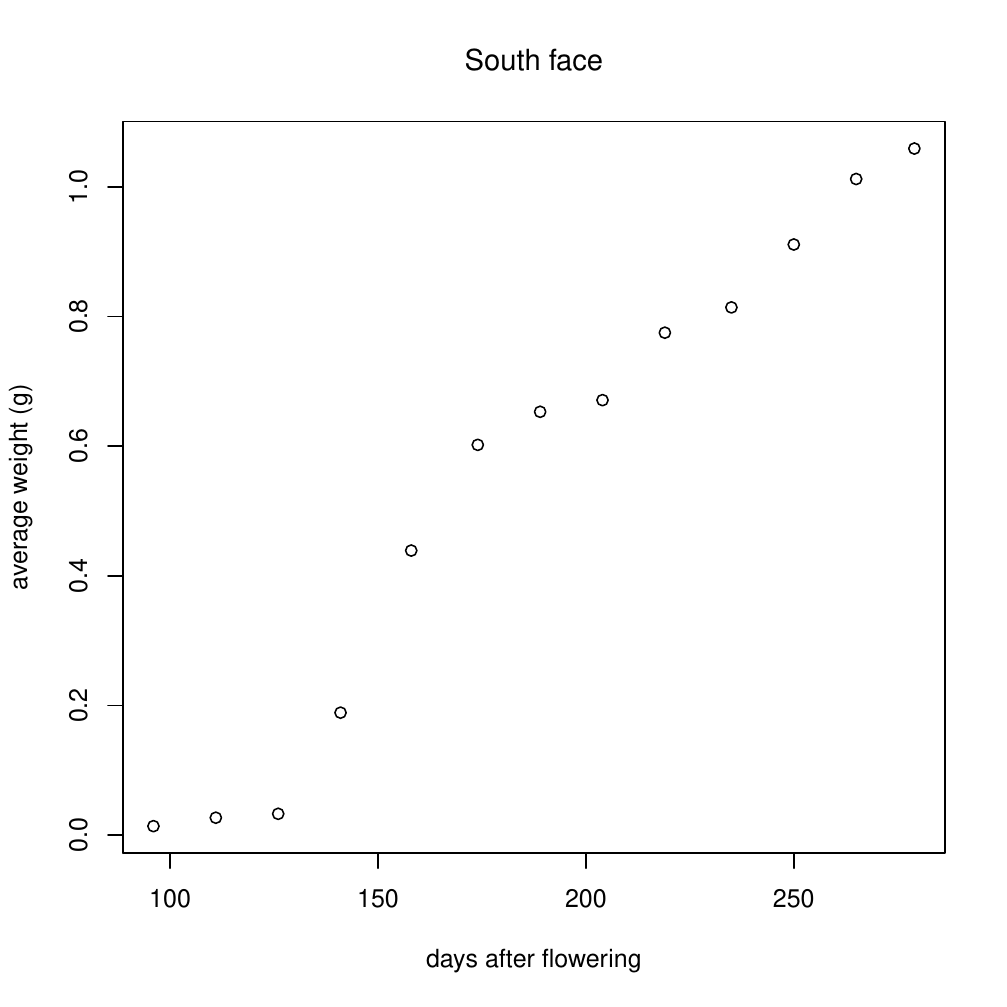}}
	\caption{Average of weight of fresh coffee berries regarding (a) North face and (b) South face.}
	\label{fig:Figura8}
\end{figure}
In order to avoid numerical problems,
according to Remark \ref{rem:1} we perform a time shifting so that the first instant is  $0$.
The fit of the data by means of multi-sigmoidal logistic function is given in Figure \ref{fig:Figura9}. More in detail, we have performed an optimization method (Nelder-Mead) to minimize the function $S_p$ defined as follows
\begin{equation*}
S_p(\theta)=\sum_{i=1}^{n}(y_i-l_m(t_i))^2,\qquad \theta=\left(\eta,\beta^T\right)^T
\end{equation*}
where $y_i$ are the real data, $t_i$ are the shifted time instants for $i=1,2,\dots,n$ and $p$ is the degree of the polynomial $Q_\beta$. Since the optimization method requires the assignment of an initial solution, we now illustrate the strategy used to obtain it.
More in detail, we note that from Eq.\ \eqref{lm} one has
\begin{equation*}
Q_\beta(t)+\log \eta =-\log\left(l_0 \frac{\eta + e^{-Q_\beta(t_0)}}{\eta l_m(t)}-1\right),
\end{equation*}
where the initial value is taken as the first observed value, i.e.\ $l_0=y_1$.
Hence, we fit the pairs $\left(t_i,-\log\left(\frac{y_n}{y_i}-1\right)\right)$,  $i=1,2,\dots,n-1$, by polynomial regression.
The corresponding estimated coefficients provide the initial values of the parameters $\beta_1,\dots, \beta_p$ and $\log{\eta}$.
\begin{figure}[t]
	\centering
	\hspace*{-1cm}
	\subfigure[]{\includegraphics[scale=0.4]{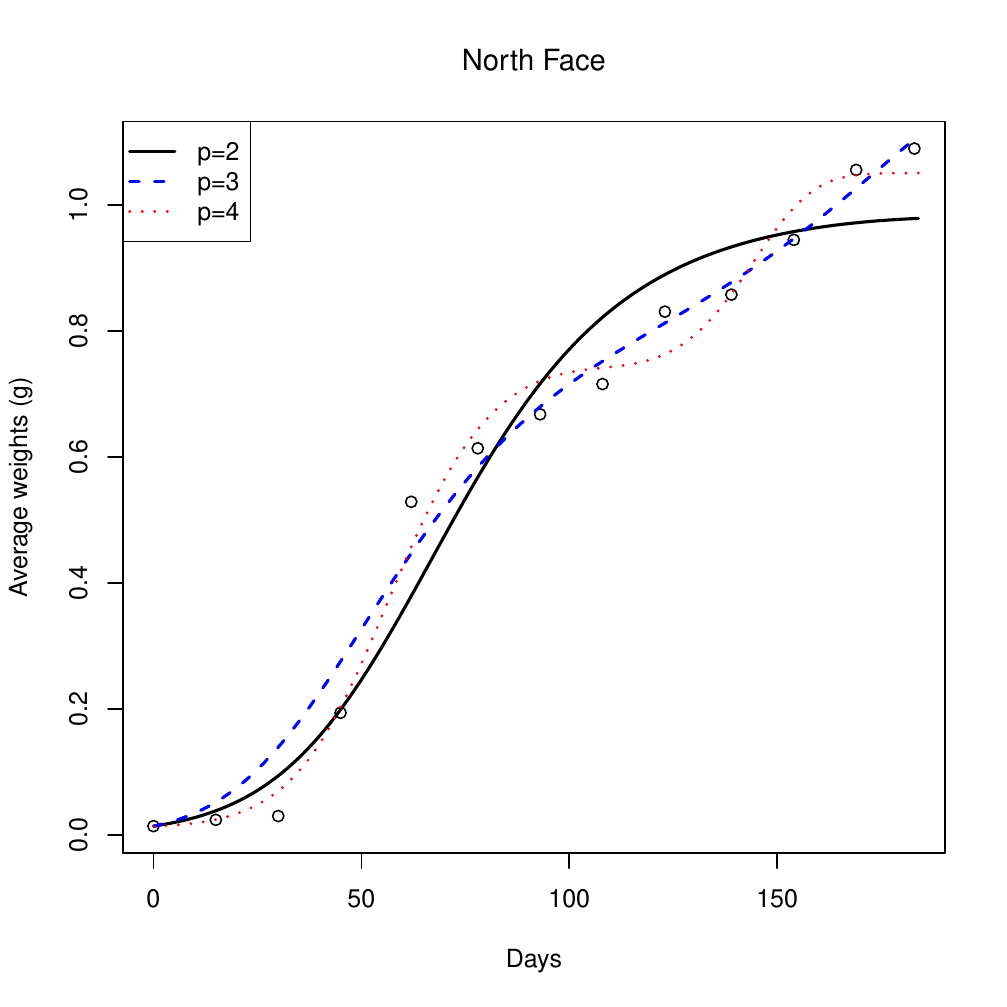}}\qquad
	\subfigure[]{\includegraphics[scale=0.4]{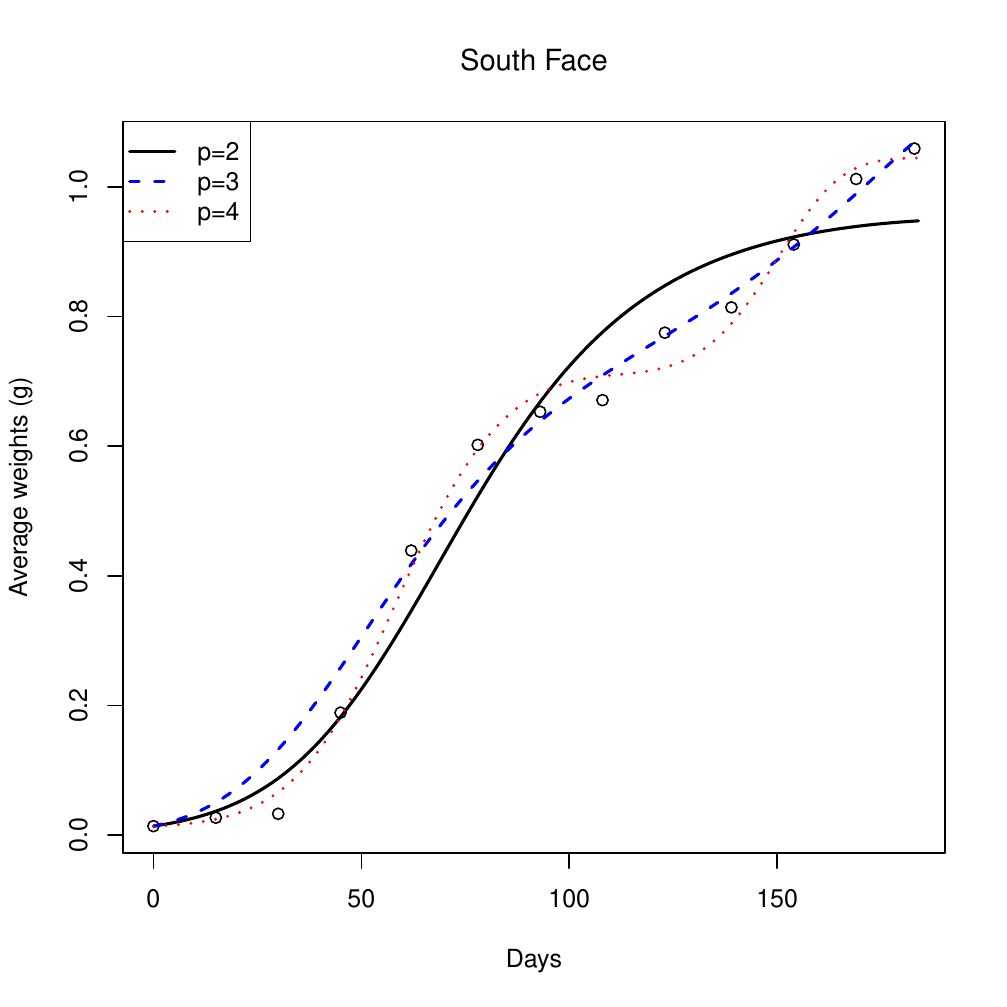}}
	\caption{Fitted multi-sigmoidal logistic curve for the coffee berries of (a) North face and (b) South face.}
	\label{fig:Figura9}
\end{figure}
%
\begin{table}[t]
	\centering
	\small
	\begin{tabular}{|c|c|c|c|c|}
		\hline
		$p$                    &$2$         &$3$               &$4$                &$4+r,\quad r\in\mathbb R$  \\ \hline
		$S_p(\tilde\theta)$ 	  &$0.07120$   &$0.03020$         &$0.01995$          &$0.01628531$ \\ \hline
		$\beta_1$     			  &$+0.07160$  &$+1.00193e-01$    &$+9.59056e-03$     &$3.617818e-02$\\ \hline
		$\beta_2$     			  &$-0.00018$  &$-7.31372e-04 $   & $+2.30393e-03$    &$1.589287e-03$ \\ \hline
		$\beta_3$     			  &-           &$+2.05499e-06$    &$-2.89567e-05$     &$-3.074182e-05$\\ \hline
		$\beta_4$                 &-           &-                 &$+1.01056e-07$     &$1.294385e-06$     \\ \hline
		$\eta$         			  &$+0.01357$  & $+1.12144e-02$   & $1.34994e-02$     &$1.316042e-02$  \\ \hline
		$r$         			  &$0$		   &$0$		          &$0$		          &$-4.498494e-01$ \\ \hline
	\end{tabular}
	\caption{The values of $S_p(\tilde\theta)$ and of the coefficients regarding coffee berries of North face.}
	\label{tab:Tabella2a}
\end{table}
\begin{table}[t]
	\centering
	\small
	\begin{tabular}{|c|c|c|c|c|}
		\hline
		$p$                    &$2$          &$3$              &$4$             &$4+r,\quad r\in\mathbb R$   \\ \hline
		{$S_p(\tilde\theta)$} 	  &$0.05916$    &$0.02155$        &$0.00799$       &$0.007781179$ \\ \hline
		$\beta_1$     			  &$+0.06886$   &$+9.79400e-02$   &$+1.41920e-02$  &$1.457478e-02$\\ \hline
		$\beta_2$     			  &$-0.00017$   &$-7.17050e-04$   &$+1.97302e-03$  &$2.012716e-03$  \\ \hline
		$\beta_3$     			  &-            &$+2.03178e-06$   &$-2.44525e-05$  &$-2.606065e-05$\\ \hline
		$\beta_4$     			  &-            &-                &$+8.30482e-08$  &$1.333322e-07$  \\ \hline
		$\eta$         			  &$0.01392$    &$1.15167e-02$    &$1.35809e-02$   &$1.373760e-02$\\ \hline
		$r$						  &$0$			&$0$			  &$0$			   &$-7.357789e-02$ \\ \hline
	\end{tabular}
\caption{The values of $S_p(\tilde\theta)$ and of the coefficients regarding coffee berries of South face.}
\label{Tab:Tabella2b}
\end{table}
We analyze the data given in Figure \ref{fig:Figura8} by using three different degrees of the polynomial $Q_\beta$,
i.e.\ $p=2$, $3$ and $4$. The values of $S_p(\theta)$ corresponding to the chosen degrees are provided
in Tables \ref{tab:Tabella2a}-\ref{Tab:Tabella2b}. The best fit, based on the minimization of $S_p(\theta)$, is
attained for  $p=4$.
\par
In order to improve the goodness-of-fit of the proposed model,
according to Remark \ref{rem:2}
the last term of the polynomial $Q_\beta$ can be modified in order to have a real exponent.
Until now the best fit is attained for $p=4$, so that hereafter we consider the following generalized model
\begin{equation*}
 \tilde l_m(t)=l_0 \frac {\eta+e^{\tilde Q_\beta(t_0)}} {\eta+e^{\tilde Q_\beta(t)}},\qquad t\ge t_0,
\end{equation*}
where
$$
 \tilde Q_\beta(t)=\beta_1t+\beta_2t^2+\beta_3t^3+\beta_4t^{4+r}, \qquad r\in \mathbb R.
$$
Hence, the aim is to find the best set of parameters $\tilde\theta=\left(\theta^T,r\right)^T$, i.e. the set which minimizes the function
\begin{equation*}
S_{4+r}(\tilde\theta)=\sum_{i=1}^{n}\left(y_i-\tilde l_m(t_i)\right)^2, \qquad\tilde\theta=\left(\theta^T,r\right)^T,
\end{equation*}
where $y_i$ and $t_i$, $i=1,\dots,n$, are respectively the real data and the shifted observation times,
and $l_0=y_1$.
We use again the Nelder-Mead optimization method with initial solutions given by $r=0$ and
the same choices for $\theta$ used in the case of integer exponents. The corresponding results,  given in the last column
of Tables \ref{tab:Tabella2a}-\ref{Tab:Tabella2b}, show that  in both cases the goodness-of-fit increases
since $S_4(\theta)>S_{4+r}(\tilde\theta)$. Moreover, this is confirmed by the plots
of the fitted models given in Figure \ref{fig:Figura15}.
%
	\begin{figure}[t]
		\centering
		\hspace*{-1cm}
		\subfigure[]{\includegraphics[scale=0.4]{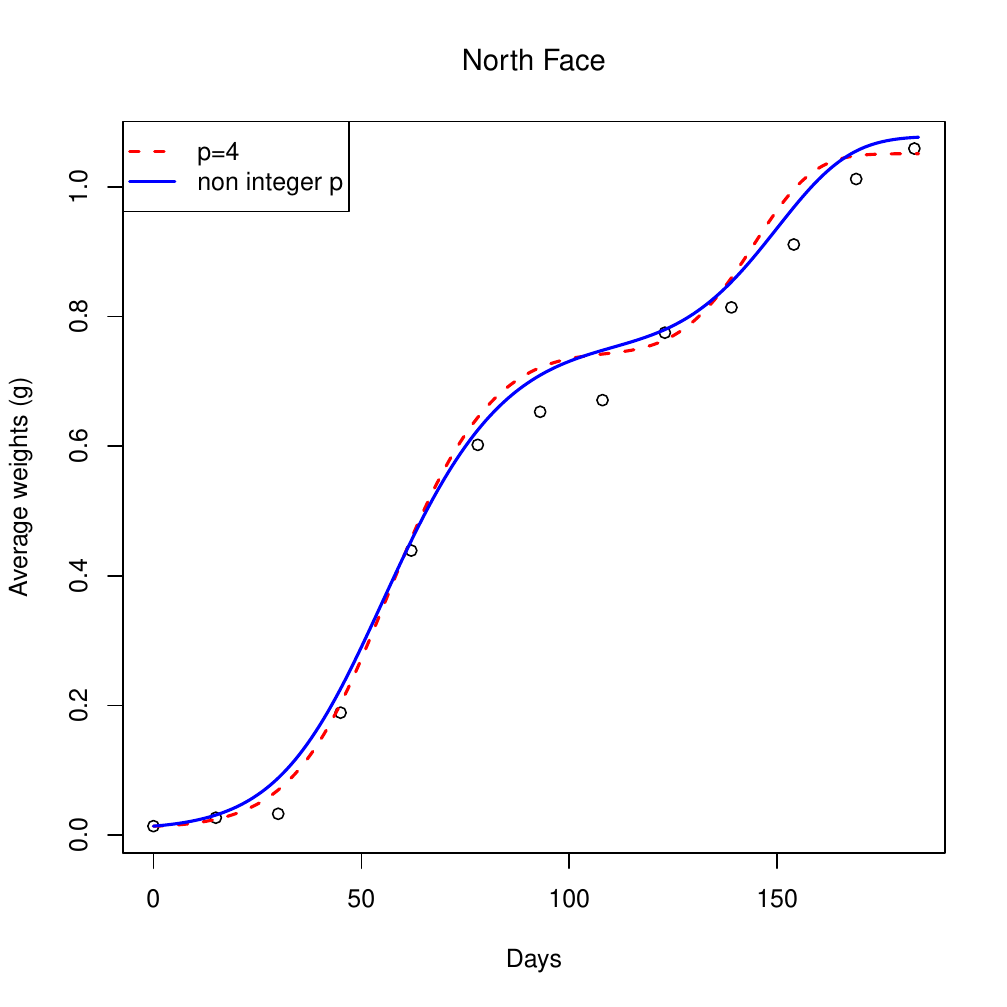}}\qquad
		\subfigure[]{\includegraphics[scale=0.4]{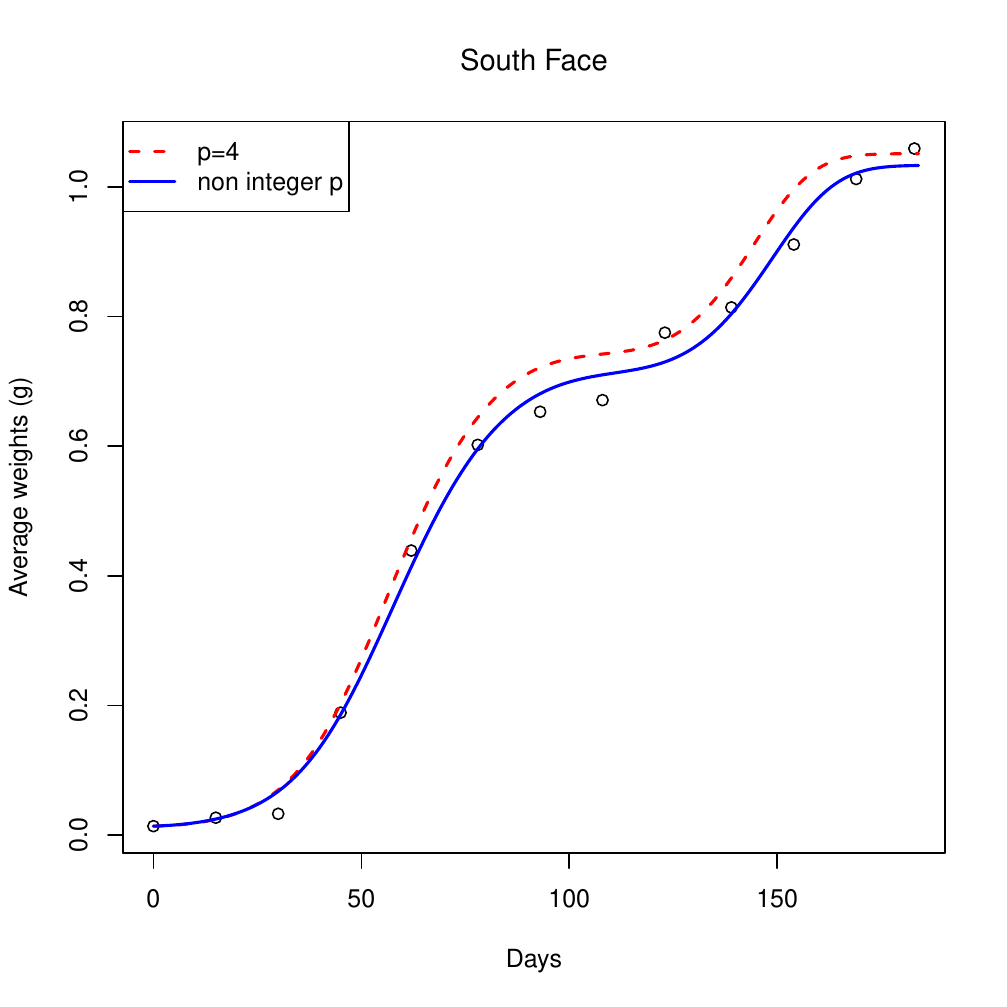}}
		\caption{Fitted generalized multi-sigmoidal logistic model for coffee berries of (a) North and (b) South face.}
		\label{fig:Figura15}
	\end{figure}	
%
\section{Analysis of a linear birth-death process}\label{section4}
Birth-death processes are often adopted to describe stochastic dynamics in various fields of biomathematics,
in ecology, genetics, and evolution. Indeed, they are appropriate to model the random evolution of the number of particles
or individuals in a system. In many cases a complete description of the probability law of such processes is not easy obtainable,
and thus one is forced to resort to computational methodologies (see, for instance, the contributions by
Crawford and Suchard \cite{CrawfordSuchard2012} and Ho {\em et al.}\ \cite{Hoetal2018}).
In this section, in order to incorporate random influences in the model described by Eq.\ \eqref{5}, we
introduce a special time-inhomogeneous birth-death process whose conditional mean is of multi-sigmoidal logistic type,
following the same strategy of  \cite{DiCrescenzoParaggio2019} and  \cite{DiCrescenzoSpina2016}.
The advantage of this approach is that the birth and death rates are linear in the size of the population
and are time-dependent in the coefficients. This allows to obtain the mean of the process in closed form,
as well as the variance in special instances.
\par
Specifically, we consider a time-inhomogeneous birth-death (BD) process $\left\{N(t);t\ge 0\right\}$
with state space $\mathbb N_0$ and linear birth and death rates given respectively by
\begin{equation}\label{3.1}
\begin{aligned}
b_n(t)&=n\lambda(t),\qquad n\in\mathbb N_0,\\
d_n(t)&=n\mu(t),\qquad n\in\mathbb N,\qquad d_0(t)=0,
\end{aligned}
\end{equation}
where the individual birth and death rates $\lambda$ and $\mu$ are integrable and positive functions
in any set $(0,t)$  with $t\ge 0$, $\mathbb N$ denotes the set of the positive natural numbers, whereas $\mathbb N_0=\mathbb N\cup \{0\}$.
Denoting by $p_n(t)=\mathbb P\left[N(t)=n\right]$ the probability that the process is in the state $n$ at the time $t$, one has that
\begin{equation}\label{2.1}
\begin{aligned}
\frac{d}{dt}p_n(t)&=d_{n+1}(t)p_{n+1}(t)+b_{n-1}(t)p_{n-1}(t)-(b_n(t)+d_n(t))p_n(t),\qquad n\in\mathbb N\\
\frac{d}{dt}p_0(t)&=d_1(t)p_1(t)-b_0(t)p_0(t).
\end{aligned}
\end{equation}
The equations of the system \eqref{2.1} are known as Chapman-Kolmogorov equations (see, for instance   \cite{Hoetal2018}).
As in several previous works, such as  in \cite{DiCrescenzoSpina2016}, we develop a probability generating function approach.  Indeed, assuming that $\mathbb P\left[N(0)=n_0\right]=1$ with $n_0\in\mathbb N$, we consider the probability generating function
$$
 G(z,t)=\mathbb E[z^{[{N(t)}|N(0)=n_0]}]= \sum_{n=0}^{\infty} \mathbb P\left[N(t)=n|N(0)=n_0\right]z^n,
 \qquad 0<z<1,\quad t> 0,
$$
with initial condition $G(z,0)=z^{n_0}$. Using a result proved in Tan \cite{Tan1986}, one has that
\begin{equation*}
G(z,t)=\left\{1-(z-1)\left[(z-1)\phi(t)-\psi(t)\right]^{-1}\right\}^{n_0},
\end{equation*}
where
\begin{equation}\label{25}
\psi(t)=\exp\left\{-\int_0^t\left[\lambda(\tau)-\mu(\tau)\right]d\tau\right\},\qquad \phi(t)
=\int_0^t \lambda(\tau)\psi(\tau)d\tau.
\end{equation}
Let us now denote the conditional mean and conditional variance by $E(t)=\mathbb E\left[X(t)|X(0)=n_0\right]$
and $Var(t)=Var\left[X(t)|X(0)=n_0\right]$, respectively. From basic properties of a linear birth-death process, we have that the conditional mean function satisfies a generalization of the classical Malthusian differential equation which is known also as exponential differential equation (see, for instance, \cite{DiCrescenzoSpina2016}), given by
\begin{equation}\label{3.2}
\frac{d}{dt} E(t)=\xi(t)E(t),\qquad t\geq 0,
\end{equation}
where $\xi$ is the net growth rate of the process. It is defined as the difference between the birth and death rate pro capite, i.e.
\begin{equation}\label{A}
\xi(t)=\lambda(t)-\mu(t).
\end{equation}
 It is easy to point out that the ODE \eqref{3.2} is formally identical to the multi-sigmoidal logistic equation \eqref{5}. Taking into account this analogy between the two aforementioned equations, similarly as in Proposition 2 of \cite{DiCrescenzoSpina2016}, one can obtain the following
\begin{proposition}\label{prop1}
	The linear birth-death process $N(t)$ with rates specified by \eqref{3.1} has conditional mean of multi-sigmoidal logistic type if, and only if, the net growth rate \eqref{A} is given by
	\begin{equation}\label{3.3}
	\xi(t)=h_\theta(t),\qquad t\ge 0
	\end{equation}
	where $h_\theta(t)$ is defined in \eqref{htheta}.
\end{proposition}
From the previous result, we have that the conditional mean of the birth-death process and the multi-sigmoidal logistic function \eqref{lm} are governed by the same ODE when the assumption \eqref{3.3} holds.
Some plots of the expected value $E_y(t)$ are provided in Figure \ref{fig:Figura11} for some choices of the parameters.
\begin{figure}[t]
	\centering
	\hspace*{-1cm}
	\subfigure[]{\includegraphics[scale=0.5]{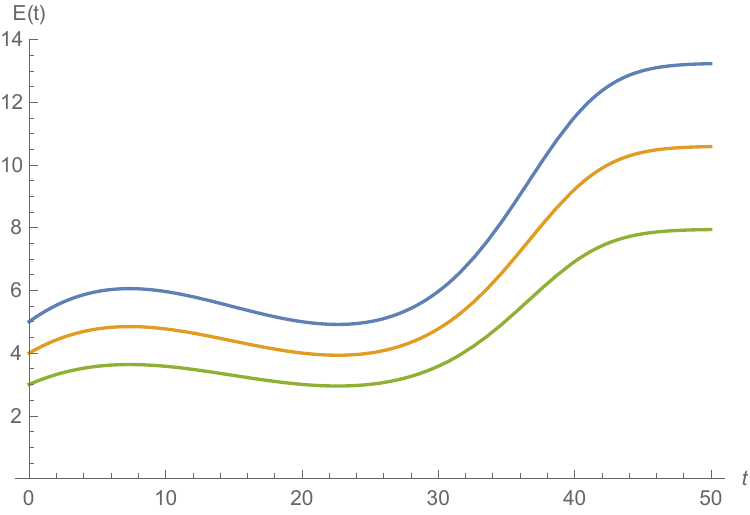}}\qquad
	\subfigure[]{\includegraphics[scale=0.5]{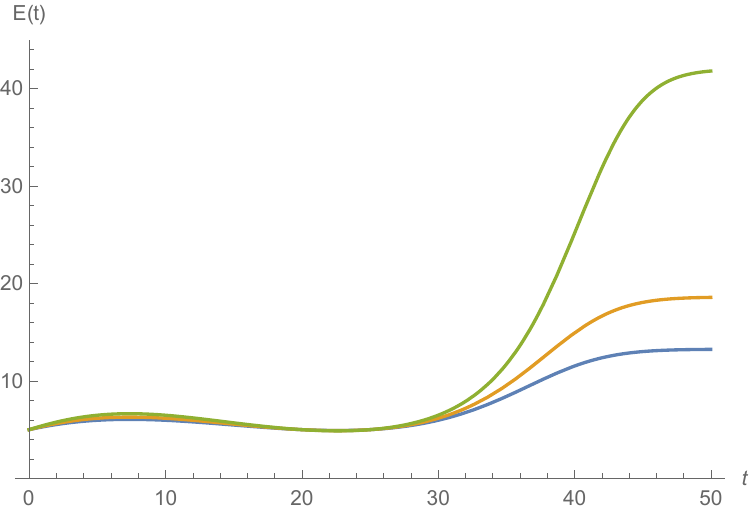}}
	\caption{The conditional mean  $E(t)$ for $Q_\beta(t)=0.1t-0.009t^2+0.0002t^3$ and
	(a) $\eta=e^{-0.5}$, $n_0=5$, $n_0=4$, $n_0=3$ (from top to bottom);
	(b) $n_0=5$, $\eta=e^{-0.5}$, $e^{-1}$, $e^{-2}$ (from bottom  to top for large $t$).}
	\label{fig:Figura11}
\end{figure}

\begin{example}\label{example}\rm
	Let the net growth rate be given as in \eqref{3.3}. We consider two suitable choices for
	$\lambda$. They are not linked together, but allow us to find a manageable expression of the variance.
	\par
	\begin{itemize}
		\item[(a)] Let  $\lambda(t)=\frac{A P_\beta(t)e^{-Q_\beta(t)}}{\eta+e^{-Q_\beta(t)}}$, with $A\ge 0$. With this assumption, the ratio between the birth and death rates is a positive constant. Considering that the conditional variance of the process is given by (cf.\ Proposition 2 of \cite{DiCrescenzoSpina2016})
		\begin{equation*}
		Var(t)=n_0 \frac{\psi(t)+2\phi(t)-1}{\psi^2(t)},
		\end{equation*}
		with  $\psi(t)$ and $\phi(t)$ specified in \eqref{25}, we have
		\begin{equation*}
		\phi(t)=A\frac{1-e^{-Q_\beta(t)}}{\eta+1},
		\end{equation*}
		so that
		\begin{equation*}
		Var(t)=\frac{y(\eta+1)(2A-1)\left(1-e^{-Q_\beta(t)}\right)}{\left(\eta+e^{-Q_\beta(t)}\right)^2}.
		\end{equation*}
		\item[(b)]If we assume $\lambda(t)=P_\beta(t)$, after some calculations it is easy to note that
		\begin{equation*}
		\phi(t)=\frac{\eta Q_\beta(t)+1-e^{-Q_\beta(t)}}{\eta +1},
		\end{equation*}
		from which we obtain the following expression for the variance
		\begin{equation*}
		Var(t)=n_0 (\eta+1)\frac{1-e^{-Q_\beta(t)}+2\eta Q_\beta(t)}{\left(\eta+e^{-Q_\beta(t)}\right)^2}.
		\end{equation*}
		See Figure \ref{fig:Figura6} for some plots of the conditional variance. Note that it is a monotic function.
		\begin{figure}[t]
			\centering
			\hspace*{-1cm}
			\subfigure[]{\includegraphics[scale=0.55]{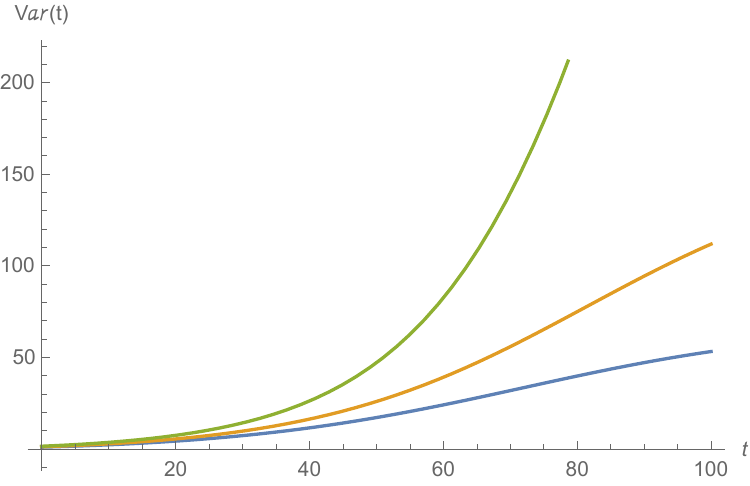}}\qquad
			\subfigure[]{\includegraphics[scale=0.55]{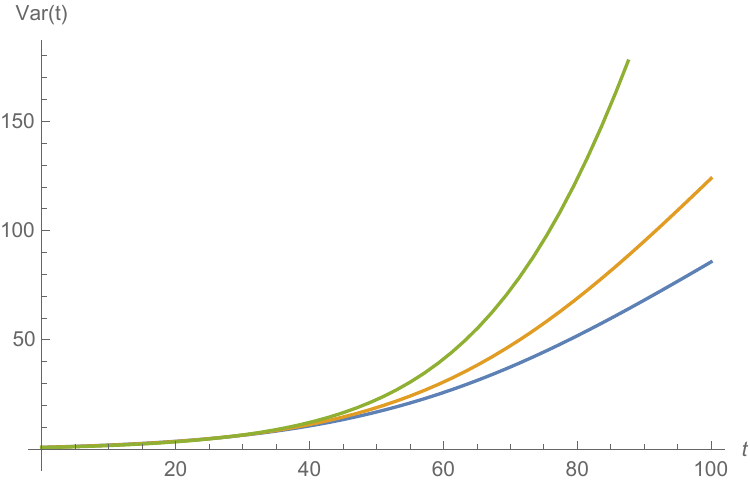}}\\
			\caption{The conditional variance $Var(t)$ in the cases (a) and (b) of Example \ref{example},
			for $n_0=5$, $Q_\beta(t)=0.1t+0.009t^2+0.0002t^3$, $A=2$ and $\eta=e^{-0.5}$, $e^{-1}$, $e^{-2}$ (from bottom  to top).}
			\label{fig:Figura6}
		\end{figure}
\end{itemize}
\end{example}
\subsection{First-passage-time problem}
The FPT problem is relevant in several applications in population dynamics, since the first
reaching of a critical high (low) level can be viewed as the rising of an overpopulation (extinction).
Hereafter we adopt an approach able to disclose the FPT densities for the birth-death process treated in this section.
\par
For a fixed threshold $n\in\mathbb N$, the FPT of the process $N(t)$ through the state $n$ starting from $N(0)=y$ is defined as follows
\begin{equation*}
T_{n_0,n}=\inf\left\{t\ge0 \colon N(t)=n\right\},\qquad N(0)=n_0.
\end{equation*}
Let us denote by  $g_{n_0,n}$  the corresponding probability density function (pdf), i.e.
\begin{equation*}
g_{n_0,n}(t)=\frac{d }{dt}P\left(T_{n_0,n}\le t\right),\qquad t\ge 0.
\end{equation*}
Considering the matrices  $A_1=\left(a^{(1)}_{i,j}\right)$ and $A_2=\left(a^{(2)}_{i,j}\right)$ defined in such a way
\begin{equation*}
a^{(1)}_{i,j}=\left\{\begin{matrix}
-i,&\quad j=i+1\\
i,&\quad j=i\\
0,&\quad \text{otherwise}
\end{matrix}\right.
\end{equation*}
for $i=1,\dots, n-2$ and
\begin{equation*}
a^{(2)}_{i,j}=\left\{\begin{matrix}
-i,&\quad j=i-1\\
i,&\quad j=i\\
0,&\quad \text{otherwise}
\end{matrix}\right.
\end{equation*}
for $i=2,\dots,n-1$, the function $g_n:=\left[g_{1,n},\dots, g_{n-1,n}\right]^T$ can be expressed as follows (cf.\ Section 3 of  \cite{Tan1986})
\begin{equation}\label{matriceG}
g_n(t)=\lambda(t)\exp\left\{-\left[A_1\Lambda(t)+A_2M(t)\right]\right\}A_1 \cdot \mathbb I_{n-1,1},
\end{equation}
where $\Lambda(t)=\int_{0}^t \lambda(\tau)d\tau$, $M(t)=\int_{0}^t \mu(\tau)d\tau$ and $\mathbb I_{n-1,1}$ is a column of all $1$ of dimension $n-1$.
Clearly, any row $g_{n_0,n}(t)$ of the vector \eqref{matriceG} is dependent from the initial state $n_0\in\left\{1,\dots,n-1\right\}$ of the process $N(t)$.
\par
It is easy to note that both $A_1$ and $A_2$ are diagonalizable, more in detail $$A_1=P_1DP_1^{-1},\qquad A_2=P_2DP_2^{-1},$$ where
\begin{equation*}
D=\diag (1,\dots, n-1),
\end{equation*}
and for any $j\in\left\{1,\dots, n-1\right\}$, the $j-th$ column of the matrix $P_1$, namely $P_1^{(j)}$ has entries given by
\begin{equation*}
x^{(1)}_{i,j}=\left\{\begin{matrix}
(-1)^m\binom{j-1}{m},&\quad i=j-m\\
0,&\quad \text{otherwise}
\end{matrix}\right.
\end{equation*}
for any $m\in\left\{0,1,\dots, j-1\right\}$ and the $j-th$ column of the matrix $P_2$, i.e.\ $P_2^{(j)}$ has entries given by
\begin{equation*}
x^{(2)}_{i,j}=\left\{\begin{matrix}
\binom{j+m}{m},&\quad i=j+m\\
0,&\quad \text{otherwise}
\end{matrix}\right.
\end{equation*}
for any $m\in\left\{0,1,\dots, n-j-1\right\}$.
\par
Therefore, taking into account that if $A=PDP^{-1}$ with $D$ diagonal, then $e^A=Pe^DP^{-1}$,
from (\ref{matriceG}) one has
\begin{equation*}
g_n(t)=\lambda(t)\left(P_1e^DP_1^{-1}\right)^{-\Lambda(t)}\left(P_2e^DP_2^{-1}\right)^{-M(t)}P_1DP_1^{-1}\mathbb I_{n-1,1}.
\end{equation*}
The latter formula provides a matrix-form expression for the FPT pdf of $N(t)$ that is computationally effective.
Some plots of the FPT pdf through $n$ are provided in Figure \ref{fig:Figura14}.
\begin{figure}[t]
	\centering
	\hspace*{-1cm}
	\subfigure{\includegraphics[scale=0.4]{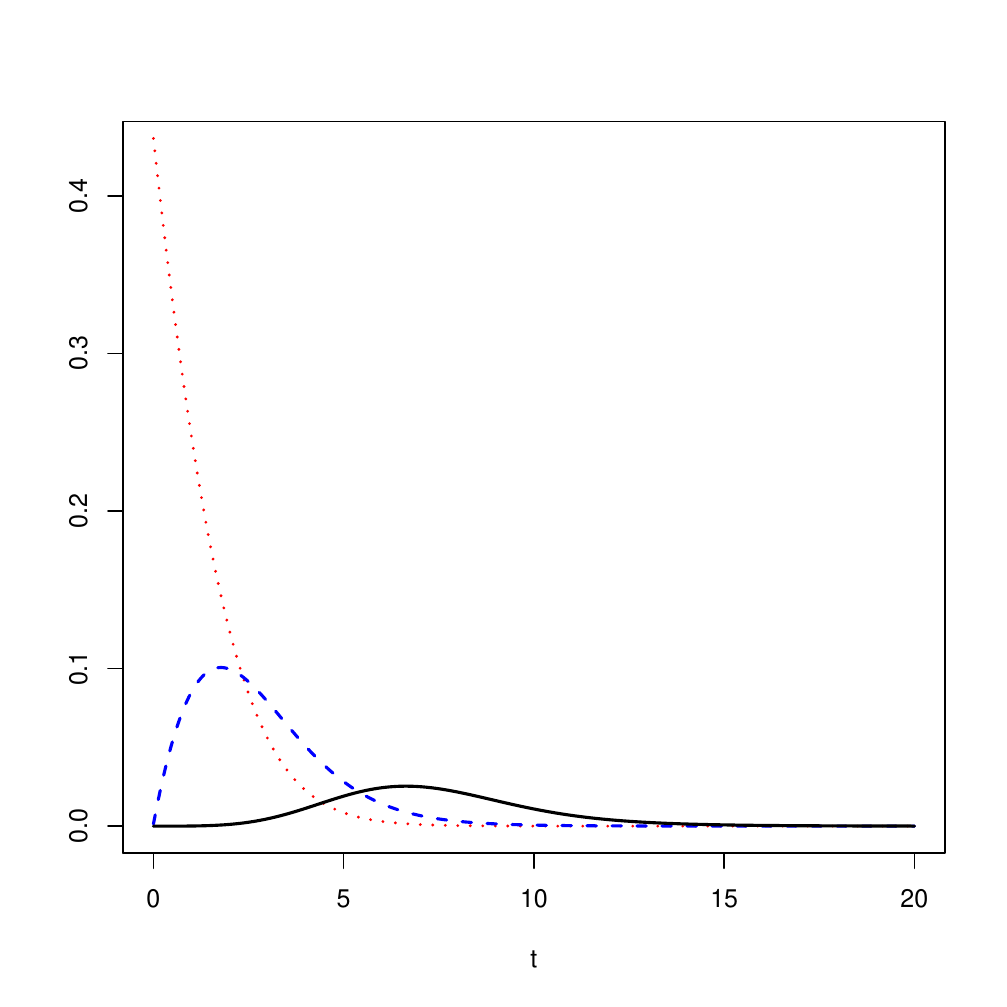}}\qquad
	\subfigure{\includegraphics[scale=0.4]{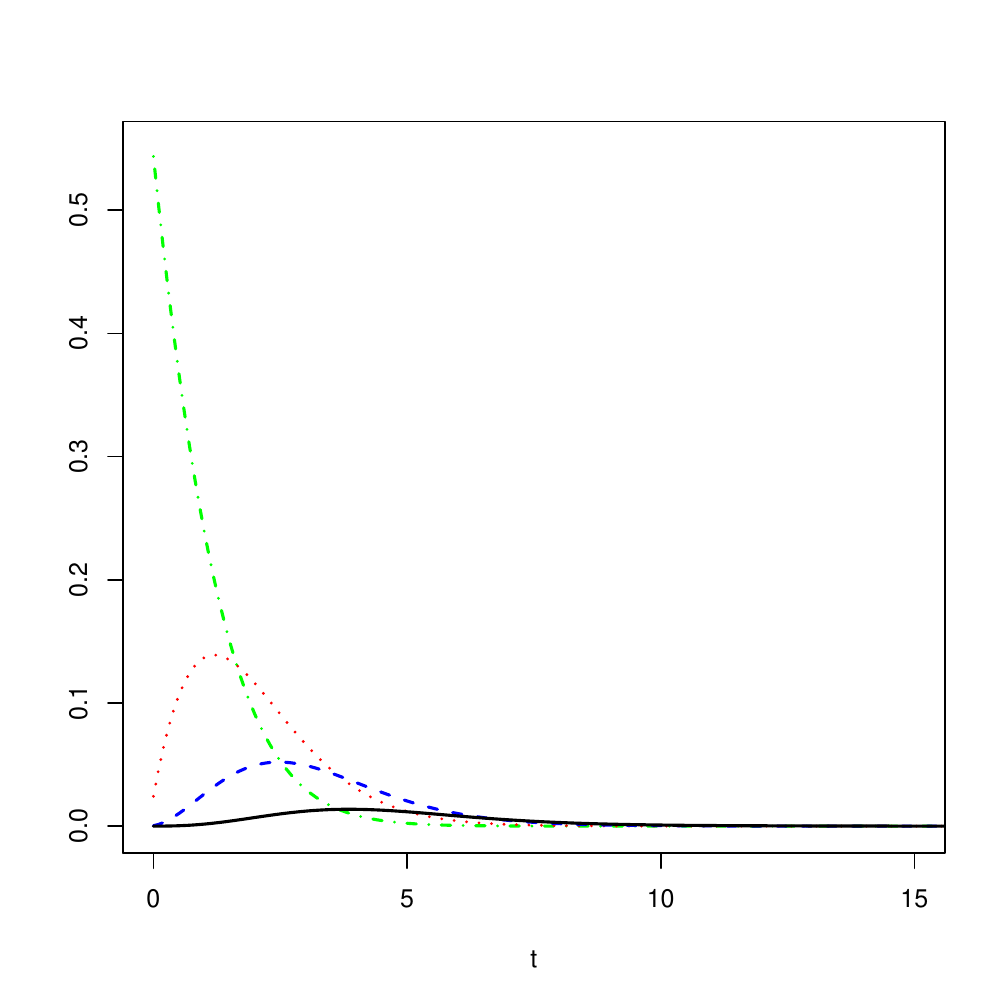}}\qquad
	\caption{The FPT pdf for $\lambda(t)=2h_\theta(t)$, $\mu(t)=h_\theta(t)$, $Q_\beta(t)=0.1t+0.009t^2+0.0002t^3$,
	$n_0=1$ (solid), $2$ (dashed), $3$ (dotted), $4$ (dot-dashed) and (a) $n=4$, (b) $n=5$.}
	\label{fig:Figura14}
\end{figure}
\section{A non linear birth-death process}\label{Section5}
In several applications in biomathematics the  systems under investigation are subject to dynamics regulated
by  linear transitions where rates are allowed to be nonlinear. Various examples emerges from  the analysis of
one-dimensional birth-death processes with quadratic rates or from two-dimensional processes with rates allowing
interaction between the components of the process (see, for instance,  \cite{Hoetal2018}).
The analysis of cases that are not solvable in closed form can be performed by adopting numerical approximation
of the transitions probabilities. A different approach is based on a suitable scaling
and limiting procedure that leads to continuous approximating processes, namely diffusion processes.
\par
Along this line, in this section we start from a special time-inhomogeneous
birth-death process having quadratic rates. The analysis is first  centered on the determination of the mean of the process,
which is multi-sigmoidal logistic. Then we obtain the asymptotic distribution of the process in terms of the Gauss hypergeometric function.
Finally, we perform a diffusive approximation leading to a non-homogeneous
lognormal diffusion process with mean of multi-sigmoidal logistic type, that will be analyzed in the next section.
\par
Let $\left\{N(t);t\ge 0\right\}$ be an inhomogeneous non-linear BD process having $\mathbb N_0$ as state space and birth and death rates given by
\begin{equation}\label{2.2}
\begin{aligned}
b_n(t)&=\lambda_1(t)+\lambda_2(t) n +\lambda_3(t) n^2,\qquad n\in\mathbb N_0,\\
d_n(t)&=\mu_1(t)+\mu_2(t) n +\mu_3(t) n^2,\qquad n\in \mathbb N,\qquad d_0(t)=0,\\
\end{aligned}
\end{equation}
where  $\lambda_1$  and  $\mu_1$  are non-negative and integrable functions and  $\lambda_i$  and  $\mu_i$  for $i=2,3$ are positive and integrable functions on any set $(0,t)$. Note that the state $0$ can be an absorbing or a reflecting endpoint.
\par
Clearly, the Eqs.\ \eqref{2.1} hold also in this special case in which the functions $b_n$ and $d_n$ are expressed by \eqref{2.2}.
%
In \cite{Valent1996},  Ismail {\em et al.}\ \cite{IsmailLetessierValent1989}  and Van Assche {\em et al.}\ \cite{VanAsscheParathasarathyLenin1999}, it is shown that the probability function $p_n$ has a particular spectral representation in terms of orthogonal polynomials also in the case of quadratic rates.
\par
With the aim of determining the mean of the process, for $0<z<1$ and $t\ge 0$, we consider the probability generating function $G(z,t)=\sum_{n=0}^{\infty}z^np_n(t)$, with initial condition $G(z,0)=z^{n_0}$ where $n_0\in\mathbb{N}_0$ is such that
\begin{equation*}
p_{n}(0)=\delta_{n,n_0}=\left\{\begin{matrix}
1,\qquad n=n_0\\
0, \qquad n\neq n_0\textcolor{blue}{.}
\end{matrix}\right.
\end{equation*}
Taking into account Eqs.\ \eqref{2.1} the probability generating function $G$ needs to verify the following PDE
\begin{equation}\label{PDE}
\begin{aligned}
\frac{\partial }{\partial t}G(z,t)&=\left(1-z\right)\left[-\frac{\mu_1(t)p_0(t)}{z}+L\left(z,\frac{\partial}{\partial z}\right)G(z,t)\right],
\end{aligned}
\end{equation}
where $L\left(z,\frac{\partial }{\partial z}\right)$ is a functional operator defined as follows:
\begin{equation*}
L\left(z,\frac{\partial}{\partial z}\right)=\frac{\mu_1(t)-\lambda_1(t) z}{z}+\left(\mu_2(t)+\mu_3(t)-z(\lambda_2(t)+\lambda_3(t))\right)\frac{\partial}{\partial z}+z\left(\mu_3(t)-\lambda_3(t)z\right)\frac{\partial^2}{\partial z^2}.
\end{equation*}
We point out that the equation \eqref{PDE} is a generalization of the one given in \cite{LetessierValent1984}, \cite{RoehnerValent1982} and  \cite{Valent1996} that can be recovered by setting $b_n=\alpha\left(n^2+bn+c\right)$ and $d_n=\alpha\left(n^2+\tilde b n\right)$.
\par
The moments of $N(t)$ are defined by
\begin{equation*}\label{defmom}
m_k(t)=\mathbb{E}[(N(t))^k|N(0)=n_0]
=\sum_{n=0}^{\infty}n^kp_n(t),\qquad  k\in\mathbb{N},
\end{equation*}
and we suppose their existence.
From Eq.\ \eqref{PDE}, by performing the derivative with respect to $z$ and taking $z\to 1$, one can easily obtain the following differential equation:
\begin{equation}\label{eqm}
\frac{d }{dt} m_1(t)=\mu_1(t)p_0(t)+\left(\lambda_1(t)-\mu_1(t)\right)+\left(\lambda_2(t)-\mu_2(t)\right)m_1(t)+\left(\lambda_3(t)-\mu_3(t)\right)m_2(t),
\end{equation}
with $m_1(0)=n_0$. If we set $\mu_1(t)=0$ and $\mu_3(t)=\lambda_3(t)$, Eq.\ \eqref{eqm} becomes
\begin{equation*}
\frac{d }{dt} m_1(t)=\lambda_1(t)+(\lambda_2(t)-\mu_2(t))m_1(t)
\end{equation*}
whose solution, taking into account the initial condition $m_1(0)=n_0$, is given by
\begin{equation*}
m_1(t)=e^{-A(t)}\left[n_0+\int_{t_0}^{t} \lambda_1(\tau)e^{A(\tau)}d\tau\right],
\end{equation*}
where $A(t)=\int_{t_0}^t (\lambda_2(\tau) -\mu_2(\tau))d\tau$.

Clearly, when $\lambda_1(t)=\mu_1(t)=0$ and $\lambda_3(t)=\mu_3(t)$, Eq.\ \eqref{eqm} can be rewritten as
\begin{equation*}
\frac{d }{dt} m_1(t)=\left(\lambda_2(t)-\mu_2(t)\right)m_1(t).
\end{equation*}
Since this is a Malthusian equation similar to Eq.\ \eqref{5}, introduced for the multi-sigmoidal logistic curve, in order to obtain a mean of multi-sigmoidal logistic type, the following condition is required: $\lambda_2(t)-\mu_2(t)=h_\theta(t)$. Indeed, with this choice, the function $m_1$, with $m_1(0)=n_0$ can be expressed as
$$
m_1(t)=n_0\frac{\eta +e^{-Q_\beta (t_0)}}{\eta + e^{-Q_\beta (t)}}.
$$
It worth noting that $m_1(t)$ is identical to the multi-sigmoidal logistic curve given in (\ref{lm})
if $l_0=n_0$ and $t_0=0$.
Note that the position $t_0=0$ in \eqref{defmom} does not affect the generality, as noted in Remark \ref{rem:1}.

\begin{remark}
		Let us assume that $\lambda_i$ and $\mu_i$ are constant, so that the rates in \eqref{2.2} are constant in time.
		Recalling the results given in  \cite{RoehnerValent1982}, the problem \eqref{2.1} admits an unique solution if the series
\begin{equation}\label{series}
	\sum_{n=0}^{\infty}\prod_{i=0}^{n}\frac{d_i}{b_i}
\end{equation}
diverges. Using the ratio criterion, the series \eqref{series} diverges if $\mu_3>\lambda_3$, and converges if $\mu_3<\lambda_3$. In the case $\lambda_3=\mu_3$, by means of  Raabe's  test of convergence, it is easy to prove that the series \eqref{series} diverges when $\mu_2+\mu_3>\lambda_2$, and converges when $\mu_2+\mu_3<\lambda_2$.
Indeed, for $a_n=\prod_{i=0}^n \frac{d_i}{b_i}$, by basic computations one has
			\begin{equation*}
			\begin{aligned}
			l=&\lim_{n\to \infty}n\left(\frac{a_n}{a_{n+1}}-1\right)=\lim_{n\to \infty}n\left(\frac{\prod_{i=0}^n \frac{d_i}{b_i}}{\prod_{i=0}^{n+1}\frac{d_i}{b_i}}-1\right)=\lim_{n\to \infty}n\left(\frac{b_{n+1}}{d_{n+1}}-1\right)\\
			&=\lim_{n\to \infty}n\left(\frac{\lambda_1+\lambda_2(n+1)+\lambda_3(n+1)^2}{\mu_1+\mu_2(n+1)+\mu_3(n+1)^2}-1\right)=\frac{\lambda_2-\mu_2}{\mu_3}.
			\end{aligned}
			\end{equation*}
			So, the series converges if $l>1$, that is $\mu_2+\mu_3<\lambda_2$. Further on, for $\mu_2+\mu_3=\lambda_2$, by the Bertrand's test the series converges since
			\begin{equation*}
			\lim_{n\to \infty} \left[n\left(\frac{a_n}{a_{n+1}}-1\right)-1\right]\log n =0.
			\end{equation*}
		Hence, the condition of existence and uniqueness is fulfilled when $\mu_3>\lambda_3$ or when $\mu_3=\lambda_3$ with $\mu_2+\mu_3\ge \lambda_2$.
	\end{remark}
\subsection{Asymptotic behavior}
Now, let us focus on the asymptotic behavior of the BD process $N(t)$. By setting
$q_n=\lim_{t\to+\infty}p_n(t)$ and by supposing that the functions $b_i$ and $d_i$
are constant with respect to $t$ for $i=1,2,3$, the system \eqref{2.1} becomes
\begin{equation}\label{qn_rel}
\begin{aligned}
0&=b_{n-1}q_{n-1}-\left(b_n+d_n\right)q_n+d_{n+1}q_{n+1},\qquad n\in\mathbb{N}\\
0&=-b_0q_0+d_1q_1,
\end{aligned}
\end{equation}
whose solutions are linked by the following iterative formula
$q_{n+1}=\frac{b_n}{d_{n+1}}q_n$, $n\in\mathbb{N}_0$.
Hence, by considering the potential coefficients defined by
$\pi_n=\frac{b_{n-1}}{d_{n}}\frac{b_{n-2}}{d_{n-1}}\dots\frac{b_{0}}{d_{1}}$, $n\in\mathbb N$ with $\pi_0=1$,
we finally have (see also Section 1.1 of Callaert and Keilson \cite{CallaertKeilson1973a})
\begin{equation}\label{qn_gen}
q_n=\frac{\pi_n}{\sum_{n=0}^{\infty}\pi_n},\qquad n\in\mathbb{N}_0.
\end{equation}
The numerical series below converges when $\mu_3>\lambda_3$ or when $\mu_3=\lambda_3$ with $\mu_2+\mu_3\ge \lambda_2$.
\par
Moreover, we can specify the behavior at the endpoint $+\infty$. Following the notation given in Callaert and Keilson \cite{CallaertKeilson1973b}, we consider
\begin{equation*}
\begin{aligned}
&A:=\sum_{n=0}^{\infty}\frac{1}{b_n\pi_n},\qquad &B:=\sum_{n=0}^{\infty}\pi_n=\sum_{n=0}^{\infty}\prod_{i=1}^{n}\frac{b_{i-1}}{d_i},\\ &C:=\sum_{n=0}^{\infty}\frac{1}{b_n\pi_n}\sum_{i=0}^{n}\pi_i,\qquad &D:=\sum_{n=0}^{\infty}\frac{1}{b_n\pi_n}
\sum_{i=n+1}^{\infty}\pi_i.
\end{aligned}
\end{equation*}
 By using the ratio criterion, if $\lambda_3<\mu_3$ ($\lambda_3>\mu_3$) it results $A=\infty$ ($A<\infty$), $B<\infty$ ($B=\infty$) and $D=\infty$ ($C<\infty$). Recalling the terminology introduced by Feller  (see, Feller \cite{Feller1959}), the endpoint $+\infty$ is natural non-attracting and unattainable when $\lambda_3<\mu_3$, instead it is an exit boundary (absorbing, attracting, attainable) when $\lambda_3>\mu_3$. As pointed out in Giorno and Nobile \cite{GiornoNobile2018}, the forward equation admits an unique solution when $+\infty$ is a natural boundary. For this reason, when $\lambda_3<\mu_3$ the existence and uniqueness of the solution of the forward equation is ensured.
\begin{example}
With the aim of studying a case in which $q_n$ can be obtained explicitly, let us now consider proportional birth and death rates, i.e.\ $\alpha\lambda_i=\mu_i$, for $i=1,2,3$ and $\alpha>0$,
so that $\alpha b_n= d_n$ for all $n\in \mathbb{N}_0$. Let us now consider two cases.
\par
{\em (i)} \ Let $\alpha=1$. By setting $a_n=b_nq_n$, Eqs.\ \eqref{qn_rel} become linear with general solution $a_n=s+nt$, for $s,t\in \mathbb R$.
Considering the initial conditions $a_0=b_0q_0=s$ and $a_1=s+t=b_1q_1=b_1\frac{b_0}{d_1}q_0=b_0q_0$ with $q_0=\frac{1}{\sum_{n=0}^{\infty}\pi_n}$ (as in Eq.\ \eqref{qn_gen}), one obtains $t=0$ and thus
\begin{equation}
 q_n=\frac{a_n}{b_n}=\frac{b_0q_0}{b_n}
 =\frac{\lambda_1}{b_n\sum_{n=0}^{\infty}\pi_n}.
 \label{eq:qconn}
\end{equation}
In order to determine the asymptotic distribution $q_n$, we suppose $\alpha\ge 1$ because the series $B$ converges only in this case.
Assuming that  $\Delta:=\lambda_2^2-4\lambda_1\lambda_3>0$, we denote by $a$ and $b$ the roots of $b_n$, so that
$b_n=\lambda_1 +\lambda_2 n +\lambda_3 n^2=\lambda_3(n-a)(n-b)$, with $a>b$. In this case, we have
$$
 \sum_{n=0}^{\infty} \pi_n
 =\lambda_1\sum_{n=0}^{\infty}\frac{1}{b_n}
 = \lambda_1\sum_{n=0}^{\infty}\frac{1}{\lambda_3(n-a)(n-b)}
 =\frac{\lambda_1}{\lambda_3(a-b)}\sum_{n=0}^{\infty} \left(\frac{1}{n-a}-\frac{1}{n-b}\right).
$$
The last series can be expressed in terms of the digamma function $\psi(z):=\frac{\Gamma'(z)}{\Gamma(z)}$, where
$\Gamma(z)=\int_{0}^{\infty}x^{z-1}e^{-x}dx$ is the Gamma function. Indeed, recalling the series expansion of $\psi$ given in
Eq.\ 6.3.16 of  Abramowitz and Stegun \cite{AbramowitzStegun1972} one has
$$
 \sum_{n=0}^{\infty} \pi_n
 =\frac{\lambda_1}{\lambda_3(a-b)} (\psi(-b)-\psi(-a)).
$$
Hence, from (\ref{eq:qconn}) we obtain
$$
q_n=\frac{\lambda_1}{\lambda_3(n-a)(n-b)\sum_{n=0}^{\infty} \pi_n}
=\frac{\sqrt{\lambda_2^2-4\lambda_1\lambda_3}}{\lambda_3(n-a)(n-b)\left(\psi(-b)-\psi(-a)\right)},
\qquad n\in \mathbb N_0.
$$
\par
{\em (ii)} \ Let $\alpha>1$. In this case,  with a similar reasoning  it can be shown that
$$
q_n=\frac{\lambda_1\sqrt{\lambda_2^2-4\lambda_1\lambda_3}}{\alpha^n \lambda_3^2(n-a)(n-b)
\left[-b\,{_2F_1\left(1,-a,1-a,\frac{1}{\alpha}\right)}+a\,{_2F_1}\left(1,-b,1-b,\frac{1}{\alpha}\right)\right]},
\qquad n\in \mathbb N_0,
$$
where $_2F_1$ is the Gauss hypergeometric function defined as
	\begin{equation*}
	_2F_1(x,y,z,t)=\sum_{k=0}^{\infty}\frac{(x)_k(y)_kt^k}{(z)_k k!},\qquad |t|<1
	\end{equation*}
	with $ (x)_k= x(x+1)\dots(x+k-1)$ for $k>0$ and $(x)_0=1$.
We remark that
$$
 \lim_{\alpha\to 1} {_2F_1\left(1,-a,1-a,\frac{1}{\alpha}\right)}
 =-a\sum_{k=0}^{\infty}\frac{1}{k-a},
$$	
so that it is not hard to check the correspondence between the  expressions of $q_n$ given in the two cases.
\par
	Some plots of the asymptotic distribution are given in Figure \ref{fig:Figura5} for some choices of the parameters.
	\begin{figure}[t]
		\centering
		\hspace*{-1cm}
		\subfigure[]{\includegraphics[scale=0.33]{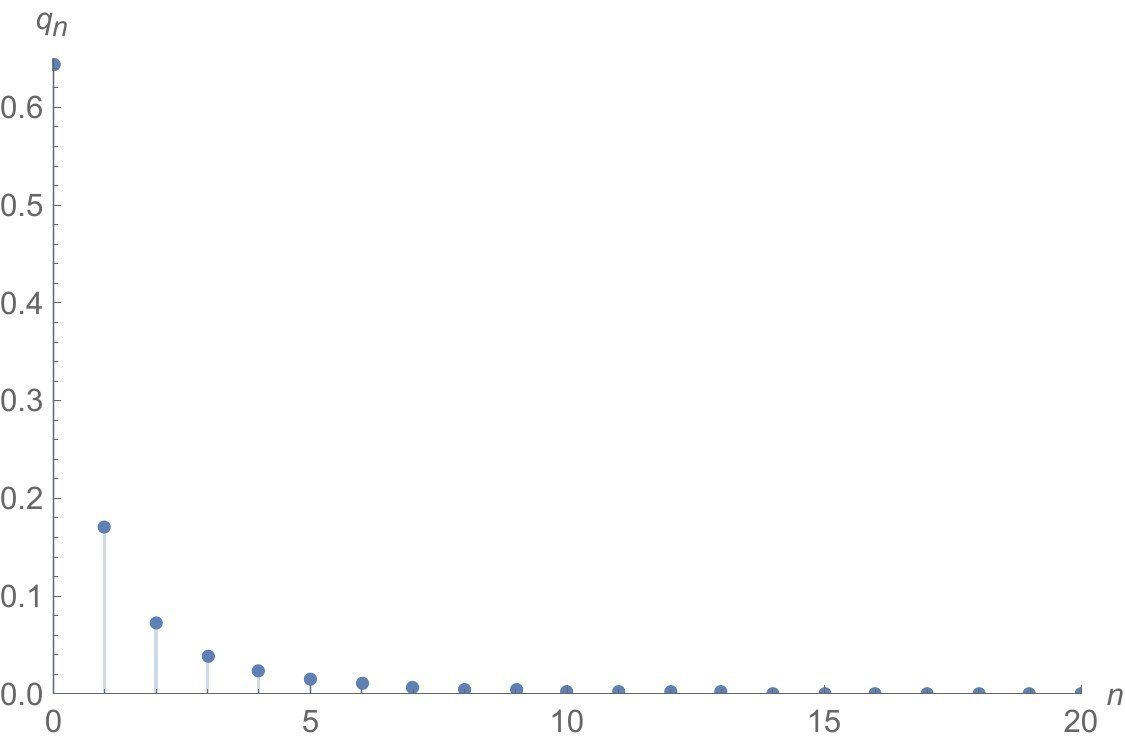}}\qquad
		\subfigure[]{\includegraphics[scale=0.33]{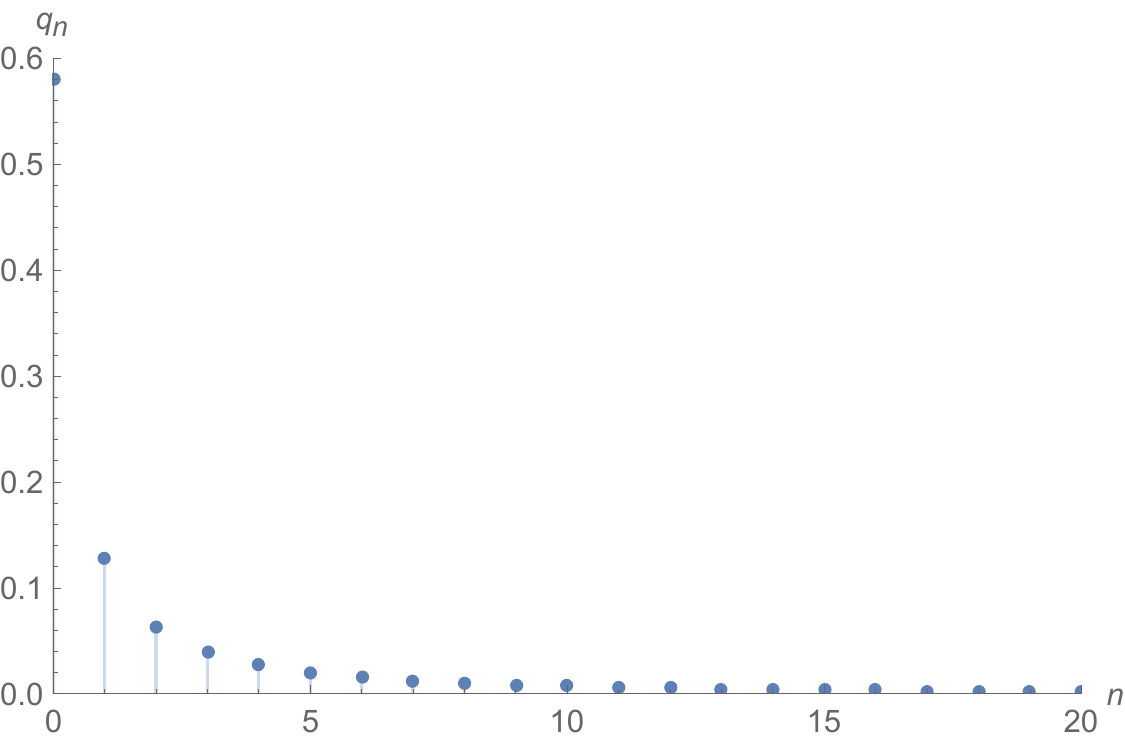}}\\
		\caption{The asymptotic distribution $q_n$ for (a) $\alpha=1.1$, $\lambda_1=1.1$, $\lambda_2=1.89$, $\lambda_3=0.8$ and (b) $\alpha=1$, $\lambda_1=2$, $\lambda_2=6.1$, $\lambda_3=1$.}
		\label{fig:Figura5}
	\end{figure}
\end{example}
	
\subsection{A diffusive approximation}
Considering the time-inhomogeneous BD process $N(t)$ with rates \eqref{2.2}, we now perform a diffusive approximation leading to a non-homogeneous lognormal diffusion process with mean of multi-sigmoidal logistic type. With the intention to obtain a more manageable description of the growth phenomenon, we introduce a suitable scaling procedure based on a scaling parameter $\varepsilon$.
More in detail,  let us consider the scaled birth-death process $N_\varepsilon(t)=\varepsilon N(t)$ whose probability $p^\varepsilon_n(t)$ solves the system \eqref{2.1}. Moreover, for $\varepsilon\simeq 0$, we have $p^\varepsilon_n(t)\simeq f(x,t)\varepsilon$ with $x=n\varepsilon$ and where  $f$ is the density function of the approximating process. The initial condition is $p^\varepsilon_{n_0}(0)=1$ with $x_0=n_0\varepsilon$.
Performing the derivative of $f$ with respect to $t$, taking into account Eqs.\ \eqref{2.1} and expanding $f$ as Taylor series around $x$, it thus results
\begin{equation*}
\begin{aligned}
\frac{\partial}{\partial t}f(x,t)&=\left[\mu_1(t)+\mu_2(t)(x+\varepsilon)+\mu_3(t)(x+\varepsilon)^2\right]\left(f(x,t)+\varepsilon\frac{\partial}{\partial x}f(x,t)+\frac{1}{2}\varepsilon^2\frac{\partial^2}{\partial x^2}f(x,t)\right)\\
&+\left[\lambda_1(t)+\lambda_2(t)(x-\varepsilon)+\lambda_3(t)(x-\varepsilon)^2\right]\left(f(x,t)-\varepsilon\frac{\partial}{\partial x}f(x,t)+\frac{1}{2}\varepsilon^2\frac{\partial^2}{\partial x^2}f(x,t)\right)\\
&-\left[\lambda_1(t)+\lambda_2(t)x+\lambda_3(t)x^2+\mu_1(t)+\mu_2(t)x+\mu_3(t)x^2\right]f(x,t)
\end{aligned}
\end{equation*}
which is equivalent to
\begin{equation}\label{fxt}
\begin{aligned}
\frac{\partial}{\partial t}f(x,t)&=\left[(\mu_2(t)-\lambda_2(t))\varepsilon +2x(\mu_3(t)-\lambda_3(t))\varepsilon + (\lambda_3(t)+\mu_3(t))\varepsilon^2\right]f(x,t)\\
&+\left[(\mu_1(t)-\lambda_1(t))\varepsilon +x(\mu_2(t)-\lambda_2(t))\varepsilon + (\lambda_2(t)+\mu_2(t))\varepsilon^2\right.\\
&+\left.x^2(\mu_3(t)-\lambda_3(t))\varepsilon +2x(\mu_3(t)+\lambda_3(t))\varepsilon^2 + (\mu_3(t)-\lambda_3(t))\varepsilon ^3\right]\frac{\partial}{\partial x}f(x,t)\\
&+\frac{1}{2}\left[(\mu_1(t)+\lambda_1(t))\varepsilon^2 +x(\mu_2(t)+\lambda_2(t))\varepsilon^2 + (\mu_2(t)-\lambda_2(t))\varepsilon^3\right.\\
&+\left.x^2(\mu_3(t)+\lambda_3(t))\varepsilon^2+2x(\mu_3(t)-\lambda_3(t))\varepsilon^3+(\lambda_3(t)+\mu_3(t))\varepsilon^4\right]\frac{\partial^2}{\partial x^2} f(x,t).
\end{aligned}
\end{equation}
We consider the positions
\begin{equation*}
\begin{array}{ll}
\lambda_1(t)=\displaystyle\frac{\alpha}{\varepsilon}+a_1(t),  & \mu_1(t)=\displaystyle\frac{\alpha}{\varepsilon}+a_2(t),\\[3mm]
\lambda_2(t)=\displaystyle\frac{\beta+r(t)}{\varepsilon}+b_1(t), & \mu_2(t)=\displaystyle\frac{\beta}{\varepsilon}+b_2(t),\\[3mm]
\lambda_3(t)=\displaystyle\frac{1}{2}\frac{\sigma^2}{\varepsilon^2}+\frac{\gamma}{\varepsilon}+c_1(t), \quad & \mu_3(t)=\displaystyle\frac{1}{2}\frac{\sigma^2}{\varepsilon^2}+\frac{\gamma}{\varepsilon}+c_2(t),
\end{array}
\end{equation*}
where $a_1$ and $a_2$ are non-negative and integrable functions, and $r$, $b_1$, $b_2$, $c_1$ and $c_2$ are
positive and integrable functions on any set $(0,t)$. Hence, the following limits hold for $\varepsilon\to 0$
\begin{equation*}
\begin{array}{lll}
(\mu_1(t)-\lambda_1(t))\varepsilon\to 0,\qquad			&(\mu_2(t)-\lambda_2(t))\varepsilon \to -r(t), \qquad 	&(\mu_3(t)-\lambda_3(t))\varepsilon\to 0\\[3mm]
(\mu_1(t)+\lambda_1(t))\varepsilon^2\to 0,\qquad		&(\mu_2(t)+\lambda_2(t))\varepsilon^2 \to 0, \qquad 			&(\mu_3(t)+\lambda_3(t))\varepsilon^2\to \sigma^2.
\end{array}
\end{equation*}
And it follows from \eqref{fxt} that  $f$  satisfies the following equation
\begin{equation*}
\frac{\partial }{\partial t}f(x,t)=-\frac{\partial }{\partial x}\left[r(t)xf(x,t)\right]+\frac{1}{2}\frac{\partial^2}{\partial x^2}\left[\sigma^2 x^2 f(x,t)\right],
\end{equation*}
which corresponds to the Fokker-Plank equation for a diffusion process $\left\{X(t); t\ge 0\right\}$ with infinitesimal moments
\begin{equation}\label{mominf}
A_1(x,t)=r(t)x,\qquad  A_2(x)=\sigma^2x^2.
\end{equation}
The initial condition $p_{n_0}(0)=1$ becomes
\begin{equation}\label{delta}
\lim_{t\to 0} f(x,t)=\delta(x-x_0),
\end{equation}
where $\delta$ is the Dirac delta function.
In Eq.\ \eqref{mominf},  we set $r(t)=h_\theta(t)$, where $h_\theta$ is defined in Eq.\ \eqref{htheta}.
Under this assumption, the process $X(t)$ will be analyzed accurately in the next section, where
in particular we show that it has  a multi-sigmoidal logistic mean.
 \section{A  diffusion process with multi-sigmoidal logistic mean}\label{Section6}
In literature, there are many stochastic differential equations used for modeling the logistic function and most of the times, although they have a solution, the resulting diffusion process is difficult to study since it is hard to find the solutions of the associated Kolmogorov equations. When the transition density function cannot be obtained, it is not useful to adopt the process for real applications.
For this reason, we address our attention to a new solvable diffusion process. More in detail, in this section we study a diffusion process whose mean is of multi-sigmoidal logistic type, following the strategy introduced in Rom\'an-Rom\'an and Torres-Ruiz \cite{RomanTorres2012,RomanTorres2015}.
 \par
 We consider a diffusion process $\left\{X(t);t\in I\right\}$ with $I=[t_0,+\infty)$ ($t_0\ge 0$), whose state space is given by $(0,+\infty)$ and whose infinitesimal moments are defined as
 \begin{equation}\label{infmom}
 A_1(x,t)=h_\theta(t)x,\qquad A_2(x)=\sigma^2x^2,
 \end{equation}
where $h_\theta(t)$ is given by \eqref{htheta}, for $\theta=\left(\eta,\beta^T\right)^T$ and $\sigma>0$.
Hence,  $X(t)$ is a lognormal diffusion  process with a time-varying drift.  Below we  show that the mean of the process is
a multi-sigmoidal logistic function.
 \par
 The process is determined by the following stochastic differential equation
 \begin{equation}\label{SDE1}
 dX(t)=h_\theta(t)X(t)dt+\sigma X(t)dW(t),\qquad X(t_0)=X_0,
 \end{equation}
 where $W(t)$ is a Wiener process, independent from the initial condition $X_0=X(t_0)$, for any $t\ge t_0$, i.e.\  a stochastic process characterized by the following properties: (i) $\mathbb P[W(t_0)=0]=1$, (ii) $W$ has independent increments, (iii) $W(t)-W(s)\sim\mathcal N(0,t-s)$. The equation \eqref{SDE1} is the stochastic counterpart of the ODE \eqref{5} and it can be easily solved by means of It\^o's formula in which we consider the variable transformation $f\left(X(t)\right)=\log\left(X(t)\right)$. In this way we obtain
 \begin{equation*}
 d\left(\log(X(t))\right)=\left(h_\theta (t)-\frac{\sigma^2}{2}\right) dt+\sigma dW(t),
 \end{equation*}
 whose solution, taking into account the initial condition $X(t_0)=X_0$ is given by
 \begin{equation*}
 X(t)=X_0\exp\left[H_\xi(t_0,t)+\sigma\left(W(t)-W(t_0)\right)\right],\qquad t\ge t_0
 \end{equation*}
where, for $t>s$,
\begin{equation*}
H_\xi(s,t)=\int_{s}^{t}h_\theta (\tau)d\tau-\frac{\sigma^2}{2}(t-s)=\log\frac{\eta+e^{-Q_\beta (s)}}{\eta+e^{-Q_\beta (t)}}-\frac{\sigma^2}{2}(t-s),
\end{equation*}
and $\xi=\left(\theta^T,\sigma^2\right)^T$.
In Figure \ref{fig:Figura7} some simulated sample paths of the multi-sigmoidal diffusion process $X(t)$ are provided.
%
\begin{figure}[t]
	\centering
	\hspace*{-1cm}
	\subfigure[]{\includegraphics[scale=0.4]{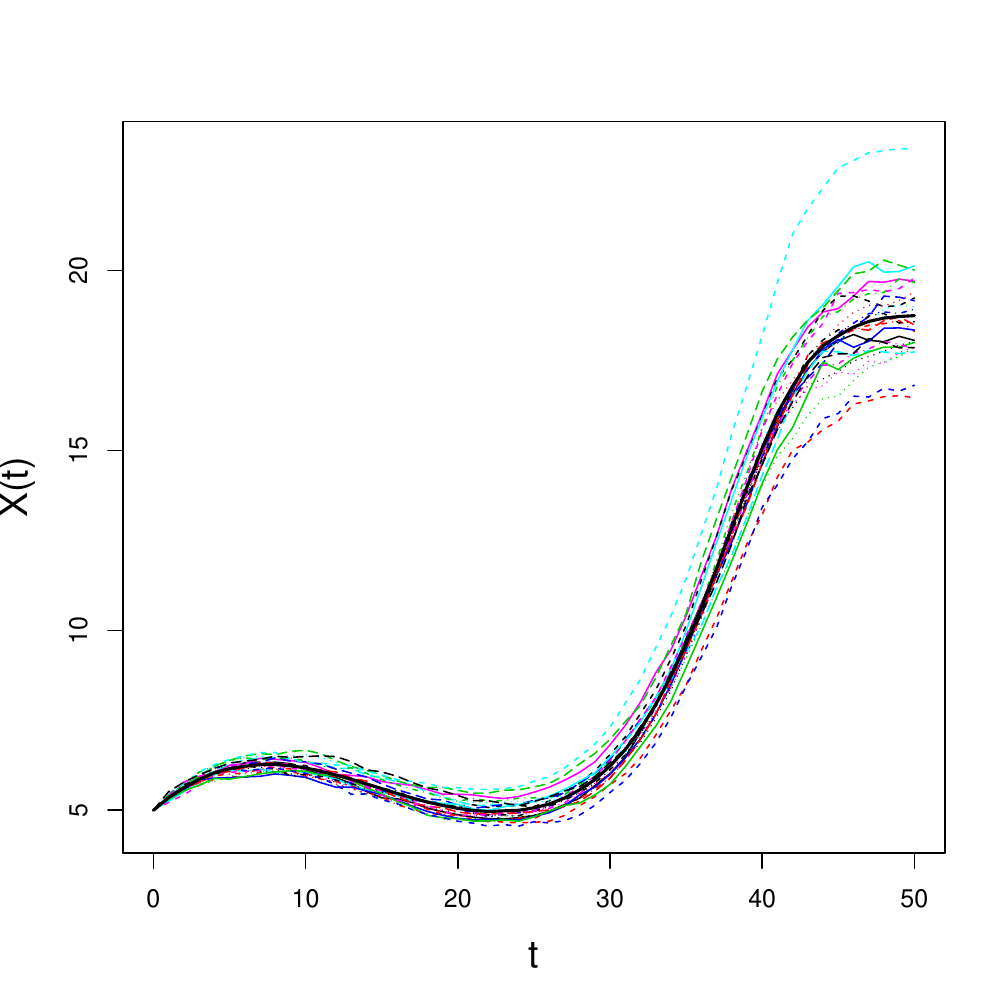}}\qquad
	\subfigure[]{\includegraphics[scale=0.4]{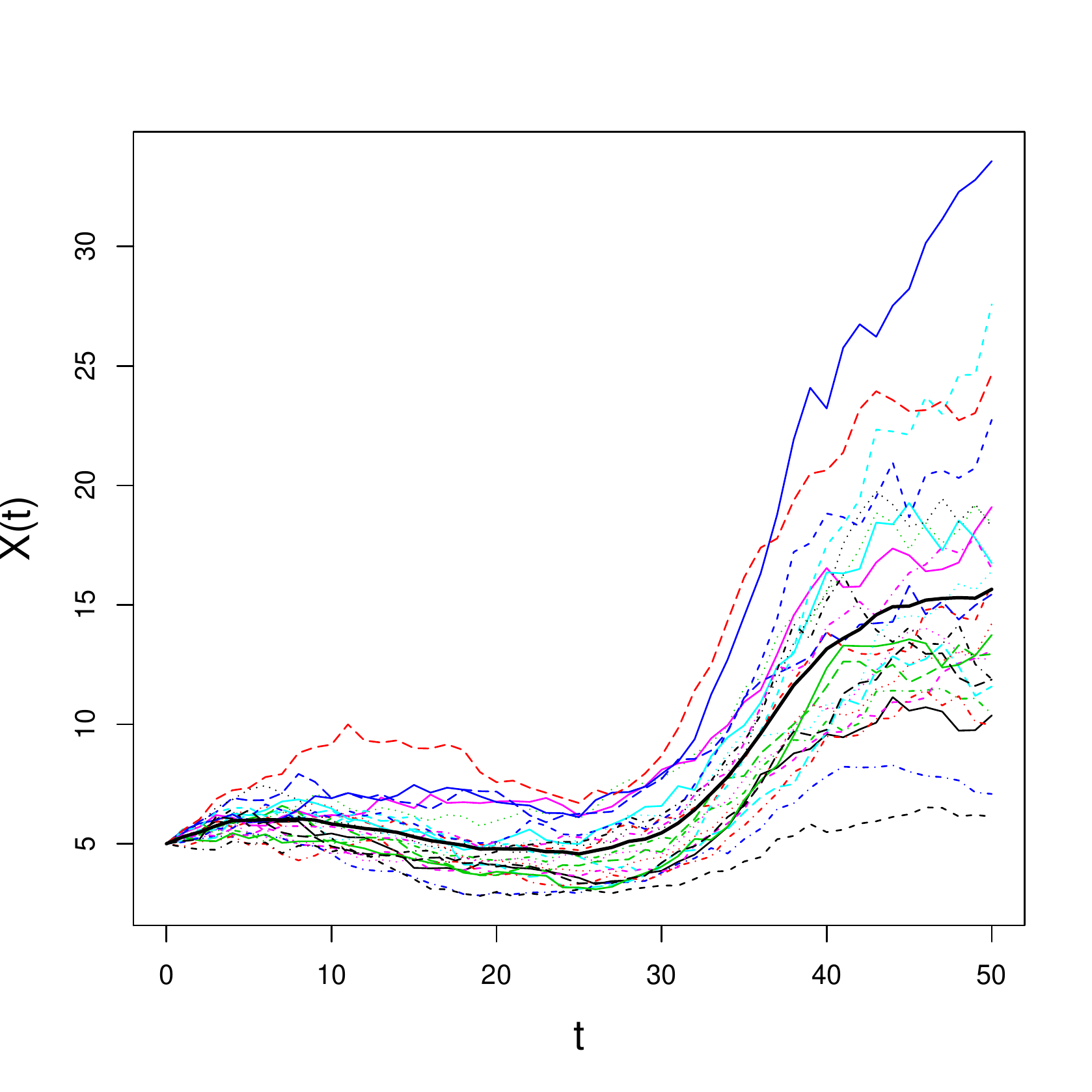}}
	\caption{Simulated sample paths of the multi-sigmoidal logistic process $X(t)$ for $x_0=5$, $\eta=e^{-1}$, $Q_\beta(t)=0.1t-0.009t^2+0.0002t^3$, (a) $\sigma=0.01$ and (b) $\sigma=0.05$. The black line represents the mean.}
	\label{fig:Figura7}
\end{figure}
\par
We can also obtain the probability distribution of the process, developing the strategy of  Rom\'an-Rom\'an and Torres-Ruiz \cite{RomanTorres2018}. More in detail, if $X_0$ is distributed according to a lognormal distribution $\Lambda_1 \left(\mu_0; \sigma_0^2\right)$, or $X_0$ is a degenerate variable (i.e.\ $ \mathbb P[X_0=x_0]=1$),
the finite dimensional distributions of the process are lognormal (note that the second case is a particular case of the former by considering $\mu_0=\ln x_0$ and $\sigma_0^2=0$). Concretely, given $n\in\mathbb{N}$ time instants  $t_1<\ldots<t_n$, the vector $(X(t_1),\ldots,X(t_n))^T$ follows an $n$-dimensional lognormal distribution $\Lambda_n(\epsilon,\Sigma)$, where the entries of the vector $\epsilon$ are given by
\begin{equation*}
\epsilon_i=\mu_0+H_\xi(t_0,t_i)=\mu_0+\log\frac{\eta+e^{-Q_\beta (t_0)}}{\eta+e^{-Q_\beta (t_i)}}-\frac{\sigma^2}{2}(t_i-t_0),
\qquad i=1,\dots,n,
\end{equation*}
and the components of the matrix $\Sigma$ are given by
\begin{equation*}
\sigma_{i,j}=\sigma_0^2+\sigma^2\left(\min\left(t_i,t_j\right)-t_0\right),\qquad i,j=1,\dots,n.
\end{equation*}
\par
Taking into account the $2$-dimensional distributions $\left(X(s), X(t)\right)^T$, with $s<t$, the transition distribution of the process is also lognormal. More in detail, we have
\begin{equation*}
\left[X(t)|X(s)=x\right] \sim\Lambda_1\left(\log x+\log\frac{\eta+e^{-Q_\beta (s)}}{\eta+e^{-Q_\beta (t)}}-\frac{\sigma^2}{2}(t-s), \sigma^2(t-s) \right),\qquad s<t.
\end{equation*}
The fact that the previous distributions are lognormal allows obtaining some of the main characteristics associated with the process.
Indeed, the $n$-th moment of $X(t)$, for $t>t_0$, is given by
\begin{equation}
\label{n-th_moment}
\mathbb{E}[X(t)^n]=\mathbb{E}[X_0^n]\left[\dfrac{\eta+e^{-Q_\beta(t_0)}}{\eta+e^{-Q_\beta(t)}}\right]^n
\exp\left(\dfrac{n(n-1)\sigma^2}{2}(t-t_0)\right), \qquad t\geq t_0
\end{equation}
whereas the $n$-th moment of $X(t)$ conditioned on $X(s)=x$ ($t_0\leq s<t$) becomes
\begin{equation}
\label{cond_n-th_moment}
\mathbb{E}[X(t)^n\mid X(s)=x]=\left[x\dfrac{\eta+e^{-Q_\beta(s)}}{\eta+e^{-Q_\beta(t)}}\right]^n
\exp\left(\dfrac{n(n-1)\sigma^2}{2}(t-s)\right), \qquad t\geq s.
\end{equation}
From \eqref{n-th_moment} and \eqref{cond_n-th_moment}, the mean and conditional mean of the process can be calculated, resulting in
\begin{equation}
\label{mean}
m(t)=\mathbb{E}[X(t)]=\mathbb{E}[X_0]\dfrac{\eta+e^{-Q_\beta(t_0)}}{\eta+e^{-Q_\beta(t)}}, \qquad t\geq t_0
\end{equation}
and
\begin{equation}
\label{cond_mean}
m(t\mid s)=\mathbb{E}[X(t)\mid X(s)=x]=x\dfrac{\eta+e^{-Q_\beta(s)}}{\eta+e^{-Q_\beta(t)}}, \qquad t\geq s.
\end{equation}
Note that the  mean \eqref{mean} and the conditional mean \eqref{cond_mean} are of multi-sigmoidal logistic type, as can been seen in  Figure \ref{fig:Figura13}, in the sense that they solve the multi-sigmoidal logistic equation \eqref{5}.
\par
Other characteristics of interest are the mode function, whose expression is given by
$$
Mode[X(t)]=Mode[X_0]\,\dfrac{\eta+e^{-Q_\beta(t_0)}}{\eta+e^{-Q_\beta(t)}}\exp\left(-\frac{3}{2}\sigma^2(t-t_0)\right), \qquad t\geq t_0,
$$
and the $\alpha$-quantile function
$$
C_\alpha[X(t)]=\dfrac{\eta+e^{-Q_\beta(t_0)}}{\eta+e^{-Q_\beta(t)}}
\exp\left(\mu_0-\dfrac{\sigma^2}{2}(t-t_0)+z_\alpha\sqrt{\sigma_0^2+\sigma^2(t-t_0)}\right), \qquad t\geq t_0
$$
from which the median function is obtained:
$$
med[X(t)]=med[X_0]\,\dfrac{\eta+e^{-Q_\beta(t_0)}}{\eta+e^{-Q_\beta(t)}}\exp\left(-\frac{\sigma^2}{2}(t-t_0)\right), \qquad t\geq t_0.
$$
We recall that $z_\alpha$ is the $\alpha$-upper quantile of a standard normal distribution. For the conditional version
of the above functions, by considering the distribution of $[X(t)\mid X(s)=x]$, $t>s$, we have
\begin{align*}
Mode[X(t)\mid X(s)=x]&=x\, \dfrac{\eta+e^{-Q_\beta(s)}}{\eta+e^{-Q_\beta(t)}}\exp\left(-\frac{3}{2}\sigma^2(t-s)\right),\\
C_\alpha[X(t)\mid X(s)=x]&=x\,\dfrac{\eta+e^{-Q_\beta(s)}}{\eta+e^{-Q_\beta(t)}}
\exp\left(-\dfrac{\sigma^2}{2}(t-s)+z_\alpha\sqrt{\sigma^2(t-s)}\right),\\
med[X(t)\mid X(s)=x]&=x\,\dfrac{\eta+e^{-Q_\beta(s)}}{\eta+e^{-Q_\beta(t)}}\exp\left(-\frac{\sigma^2}{2}(t-s)\right),
\end{align*}
respectively.
\begin{figure}[t]
	\centering
	\hspace*{-1cm}
	\subfigure[]{\includegraphics[scale=0.5]{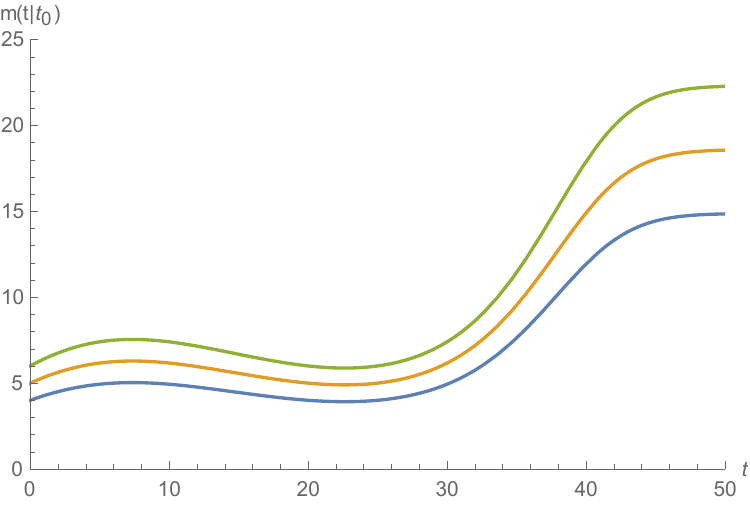}}\qquad
	\subfigure[]{\includegraphics[scale=0.5]{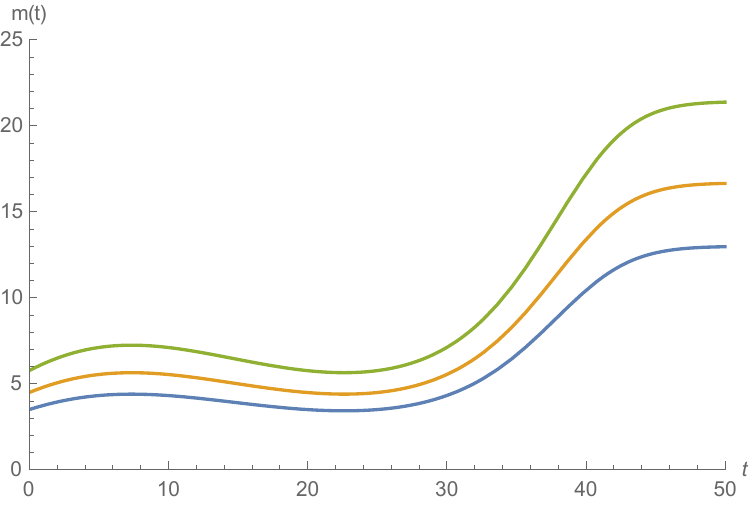}}\qquad
	\caption{For $\eta=e^{-1}$, $Q_\beta(t)=0.1t-0.009t^2+0.0002t^3$ and $t_0=0$, (a) the conditional mean $m(t|t_0)$
	with $x=4$, $5$, $6$ (from bottom to top), (b) the expected value $m(t)$ with
	$X_0\sim\Lambda_1(\mu_0, \sigma_0^2)$, for $\mu_0= 1.25$, $1.5$,  $1.75$ (from bottom to top) and   $\sigma_0^2=10^{-4}$.}
	\label{fig:Figura13}
\end{figure}
%
\subsection{First-passage-time problem}
Let us now focus on the FPT problem for the diffusion process $X(t)$ with infinitesimal moments given by \eqref{infmom} in analogy with the analysis of the threshold-crossing problem of Section \ref{TCP}.
\par
Given a continuous function $S$ defined on $I=[t_0,+\infty)$, we define the FPT of the process $X(t)$ through the boundary
$S(t),\; t\in I$, conditioned on $x_0$ as
\begin{equation*}
T_{x_0}=
\left\{
\begin{array}{ll}
 \inf\left\{t\ge t_0 \colon X(t)>S(t)\,|\, X(t_0)=x_0\right\}, &  x_0<S(t_0), \\
 \inf\left\{t\ge t_0 \colon X(t)<S(t)\,|\, X(t_0)=x_0\right\}, &  x_0>S(t_0).
\end{array}
\right.
\end{equation*}
Denoting by
\begin{equation*}
g\left(S(t), t|x_0,t_0\right)=\frac{d \mathbb P(T_{x_0}\le t)}{dt}
\end{equation*}
the corresponding pdf, it is well known that $g$ satisfies the following Volterra integral equation (cf.\  \cite{GiornoNobile2019} and \cite{Gutierrez1997})
\begin{equation}\label{eqVolt}
g\left(S(t), t|x_0,t_0\right)=\rho\left\{-q\left(S(t),t|x_0,t_0\right)+2\int_{t_0}^{t} g\left(S(\tau), \tau|x_0,t_0\right)q\left(S(t),t|S(\tau),\tau\right)d\tau\right\},\quad t\ge t_0
\end{equation}
with $\rho=\sgn\left(S(t_0)-x_0\right)$ and
\begin{equation*}
\begin{aligned}
q\left(S(t), t|x_0,\tau\right)&=\frac{1}{2}f\left(S(t),t|x_0,\tau\right)\left[S'(t)-S(t)h_\theta\left(t\right)+\frac{3}{2}\sigma^2 S(t)\right]\\
&+\frac{1}{2}\sigma^2S(t)^2\left.\frac{\partial}{\partial x}f\left(x,t|x_0,\tau\right)\right|_{x=S(t)},
\end{aligned}
\end{equation*}
where $f$ is the transition pdf of the process $X(t)$ and $h_\theta$ is defined in \eqref{htheta}.
\par
In general, the solutions of Eq.\ \eqref{eqVolt} cannot be expressed in a closed form but it is possible only in certain special cases.
For example, considering a lognormal process having constant infinitesimal moments, it is possible to determine a closed form for the FPT pdf $g$ in the presence of a constant threshold $S(t)=n$. Unfortunately, in our case study, because of the time-dependent drift, a closed form for the FPT pdf in the presence of a constant threshold cannot be obtained easily. However, in this case a suitable transformation leads to a time homogeneous process in the presence of a time-dependent threshold, for which the FPT pdf can be obtain explicitly. More in detail, taking into account the results given in the Example on p.\ 630 of   \cite{Gutierrez1997},
we can obtain a closed-form expression for the FPT pdf when
\begin{equation*}
S(t)=\exp\left[A+Bt+\int h_\theta(t)dt\right]=\frac{e^{A+Bt}}{\eta+e^{-Q_\beta(t)}},
\end{equation*}
with $A,B\in\mathbb R$. The boundary $S$ and the FPT pdf  of the process $X(t)$ through $S$ are plotted in Figure \ref{fig:Figura12}.
\begin{figure}[t]
	\centering
	\hspace*{-1cm}
	\subfigure[]{\includegraphics[scale=0.5]{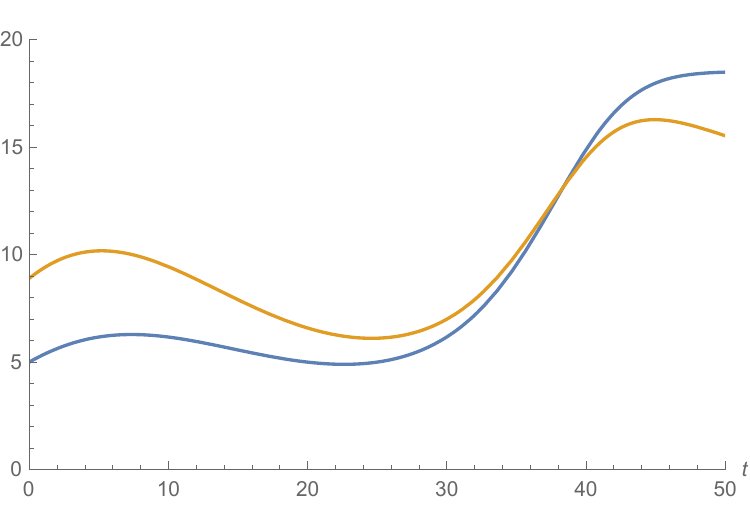}}\qquad
	\subfigure[]{\includegraphics[scale=0.5]{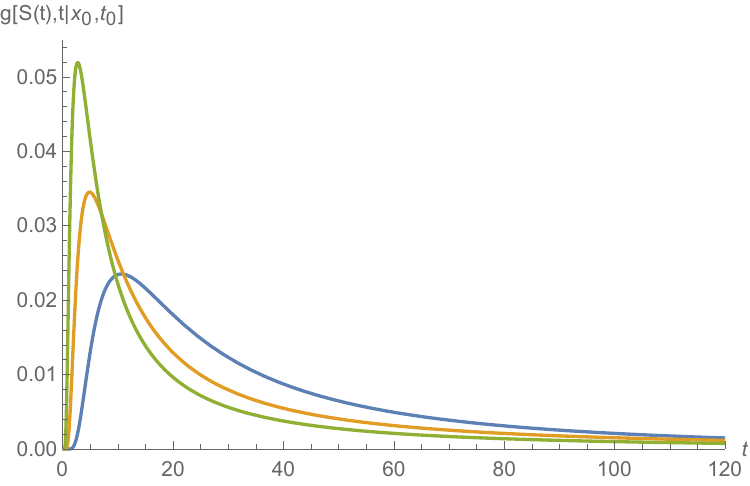}}\qquad
	\caption{For $x_0=5$, $\eta=e^{-1}$, $Q_\beta(t)=0.1t-0.009t^2+0.0002t^3$, $A=2.5$, $B=-0.015$, $t_0=0$ (a) the boundary $S(t)$ (upper curve near the origin) and the conditional expected value $m(t|t_0)$ of the process (lower curve near the origin), (b) the pdf of the FPT through $S(t)$ with $\sigma=0.1$, $0.15$, $0.2$ (from bottom to top near the origin).}
	\label{fig:Figura12}
\end{figure}
Precisely, in this case the pdf of the FPT is given by
\begin{equation*}
\begin{aligned}
g\left(S(t), t|x_0,t_0\right)&=\frac{\left|\log\left(\frac{x_0}{S(t_0)}\right)\right|}{\sqrt{2\pi\sigma^2(t-t_0)^3}}\exp\left[\frac{-\left(\log\frac{S(t)}{x_0} -\log \frac{\eta+e^{-Q_\beta(t_0)}}{\eta+e^{-Q_\beta(t)}} +\frac{\sigma^2}{2}(t-t_0)\right)^2}{2\sigma^2(t-t_0)}\right], \qquad t> t_0,
\end{aligned}
\end{equation*}
with $S(t_0)\neq x_0$.
\par
Otherwise, when $S$ is a different boundary, the reader is referred to Buonocore {\em et al.}\ \cite{BuonocoreNobileRicciardi1987} and Rom\'an-Rom\'an {\em et al.}\
\cite{RomanPerezTorres2012}, where the problem is studied by means of numerical methods.
The analysis of the first-passage-time problem by means of this numerical approach will be the object of a future investigation.
%
\section{Conclusions}
In recent years, sigmoidal curves have been used in many fields of applications and in order to include random fluctuations, typical of real world, various stochastic processes have been defined. The present paper has been devoted to a generalization of the classical logistic growth including more than one inflection point. This curve, called multi-sigmoidal logistic function, has been studied both from a deterministic and stochastic point of view. An application involving real data has been also performed to point out the usefulness of the aforementioned curve.
Further on, in order to improve the goodness-of-fit of the proposed model, we added a term with rational non-integer greater than one exponent to the polynomial. Moreover, two different birth-death processes have been introduced, with linear and quadratic birth and death rates. In both cases, we have investigated the conditions under which they present a mean of multi-sigmoidal logistic type. Finally, with the aim of obtaining a more manageable stochastic description of the growth, we have performed a suitable diffusion scaling leading to a special lognormal diffusion process
with multi-sigmoidal logistic mean. Many features of the approximating diffusion process
have been analyzed and also the FPT pdf through particular boundaries has been obtained.
Clearly, in order to use the stochastic model in real applications, an estimation study is needed.
This will be the object of the next investigation, where this kind of problem will be analyzed in detail
by means of statistical tools and numerical methods.

\subsection*{\bf Author Contributions}
All the authors contributed equally to this work.

\subsection*{\bf Acknowledgements}
\small Antonio Di Crescenzo\ and Paola Paraggio\ are members of the research group GNCS
of INdAM (Istituto Nazionale di Alta Matematica). \\
\small This work was support in part by the Ministerio de Econom\'ia, Industria y Competitividad, Spain, under Grant MTM2017-85568-P, FEDER/Junta de Andaluc\'{\i}a-Consejer\'ia de Econom\'ia y Conocimiento, under Grant A-FQM-456-UGR18 and by Italian MIUR-PRIN 2017,
project `Stochastic Models for Complex Systems', No.\ 2017JFFHSH.\\
\small Paola Paraggio\ thanks the Department of Statistics and Operations Research, Faculty of Sciences of the University of Granada and the Institute of Mathematics of the University of Granada (IEMath-GR) for the hospitality during the one-month visit carried out in 2019.

%
\end{document}